\documentclass[11pt]{article}
\usepackage{amsmath,amssymb,amsthm,latexsym}
\usepackage[dvips]{graphics}
\usepackage{epsfig, rotate}
\usepackage[english]{babel}
\usepackage{amsfonts}
\usepackage{array}
\usepackage{color}
\usepackage{url}

\newtheorem{theorem}{{Theorem}}[section]

\newtheorem{corollary}{{Corollary}}[section]
\newtheorem{lemma}{{Lemma}}[section]

\numberwithin{equation}{section} \numberwithin{theorem}{section}
\numberwithin{lemma}{section}\numberwithin{corollary}{section}

\newcommand{\be}{\begin{equation}}
\newcommand{\ee}{\end{equation}}
\newcommand{\beaa}{\begin{eqnarray*}}
\newcommand{\eeaa}{\end{eqnarray*}}
\newcommand{\bea}{\begin{eqnarray}}
\newcommand{\eea}{\end{eqnarray}}

\newcommand{\bei}{\begin{itemize}}
\newcommand{\eei}{\end{itemize}}

\numberwithin{equation}{section}

\def\E{\mathbb{E}}

\def\td{\stackrel{d}{\rightarrow}}

\begin{document}
\thispagestyle{empty}


\noindent {\bf \Large Pearson's  goodness-of-fit tests for sparse
distributions}

\vspace{20pt} \noindent{\bf Shuhua Chang$^{a,b}$, Deli Li$^c$,
Yongcheng Qi$^d$}

\vspace{10pt}

{\small \noindent $^a$Coordinated Innovation Center for Computable
Modeling in Management Science, Yango University, Fujian 350015,
China

\noindent $^b$Coordinated Innovation Center for Computable Modeling
in Management Science, Tianjin University of Finance and Economics,
Tianjin 300222, China
\\ Email: szhang@tjufe.edu.cn

\vspace{10pt}

\noindent $^c$Department of Mathematical Sciences, Lakehead
University Thunder Bay,  Ontario, Canada P7B 5E1.
\\Email: dli@lakeheadu.ca

\vspace{10pt}

\noindent $^d$Department of Mathematics and Statistics, University
of Minnesota Duluth, 1117 University Drive, Duluth, MN 55812, USA.
\\Email: yqi@d.umn.edu

\footnotetext[1]{This is an Accepted Manuscript version of the
following article, accepted for publication in {\it Journal of
Applied Statistics} [https://doi.org/10.1080/02664763.2021.2017413].
 It is deposited under the terms of the Creative
Commons Attribution-NonCommercial License
(http://creativecommons.org/licenses/by-nc/4.0/), which permits
non-commercial re-use, distribution, and reproduction in any medium,
provided the original work is properly cited.}

\date{\today}

\vspace{20pt}

\noindent{\bf Abstract.} Pearson's chi-squared test is widely used
to test the goodness of fit between categorical data and a given
discrete distribution function. When the number of sets of the
categorical data, say $k$, is a fixed integer,  Pearson's
chi-squared test statistic converges in distribution to a
chi-squared distribution with $k-1$ degrees of freedom when the
sample size $n$ goes to infinity. In real applications, the number
$k$ often changes with $n$ and may be even much larger than $n$. By
using the martingale techniques, we prove that Pearson's chi-squared
test statistic converges to the normal under quite general
conditions. We also propose a new test statistic which is more
powerful than chi-squared test statistic based on our simulation
study.  A real application to lottery data is provided to illustrate
our methodology.


\vspace{20pt}


\vspace{10pt}


\newpage

\section{Introduction}\label{introduction}
Consider an experiment which can result in $k$ possible events, say,
$E_1, \cdots, E_k$, where $k\ge 2$ is an integer and $E_1, \cdots,
E_k$ form a partition of the sample space. Repeat the experiment $n$
times independently and let $o_i$ denote the observed frequency of
event $E_i$ for $i\in \{1, \cdots, k\}$. Then $(o_1,\cdots, o_k)$
has a multinomial distribution. To test the null hypothesis that the
probabilities for events $E_1, \cdots, E_k$ are equal to $p_1,
\cdots, p_k$, respectively,  where $p_1, \cdots, p_k$ are $k$
specified positive numbers with $p_1+\cdots+p_k=1$, define the
following chi-squared test statistic
\begin{equation}\label{pearson}
\mathcal{X}^2_n=\sum^k_{i=1}\frac{(o_i-e_i)^2}{e_i},
\end{equation}
where $e_i=np_i$ is the expected number of events $E_i$ to occur in
the $n$ trials of the experiment for $i\in \{1, \cdots, k\}$. The
limiting distribution of $\mathcal{X}^2_n$ is a chi-squared
distribution with $k-1$ degrees of freedom when $k$ is a fixed
integer.  This is the well-known chi-squared goodness-of-fit test
proposed by Pearson~\cite{Pearson}. A test with approximate size
$\alpha$ rejects the null hypothesis if
$\mathcal{X}^2_n>\chi^2_{k-1}(\alpha)$, where $\chi^2_{k-1}(\alpha)$
denotes the $\alpha$-level critical value of a chi-squared
distribution with $k-1$ degrees of freedom for $\alpha\in (0,1)$.


As a statistical method, Pearson's chi-squared goodness-of-fit test is one of the most popular topics
offered in college statistics courses.  The above testing problem can be  restated in a different form.
Let $f_0$ be a discrete probability mass function defined over $\{x_i, 1\le i\le k\}$ and set $p_i=f_0(x_i)$ for $1\le i\le  k$.
Assume that a random sample of size $n$, $X_1, \cdots, X_n$, is drawn from the distribution of a discrete random variable $X$,
 where $X$ is a discrete random variable having a probability mass function $f(x)$
 for $x=x_1, \cdots, x_k$. Now we can define $E_i=\{x_i\}$ for $1\le i\le k$ and
 set $o_i=\sum^n_{j=1}I(X_j\in E_i)$ for $1\le i\le k$, and $e_i=np_i$. Then Pearson's test statistic $\mathcal{X}^2_n$ defined
in \eqref{pearson} can be used to test hypothesis $H_0:~f=f_0 $;
i.e. $f(x_i)=f_0(x_i)$ for $1\le i\le k$. Traditionally, Pearson's
chi-squared goodness-of-fit test is suggested to use only if the
value of $k$ is relatively small compared with the sample size $n$.
When there are infinite many values for a discrete random variable
$X$ or the number of distinct values of $X$ is
  too large compared with the sample size $n$,  one can first
  select a proper integer $k$ and then re-group values of $X$ into $k$ categories by putting the values of $X$  with small probabilities
  (under the null hypothesis) into one category.  When  $f_0$ is a probability density function, one can discretize the variable $X$ so that
   Pearson's chi-squared goodness-of-fit test can be used to test whether the density function of $X$ is equal to $f_0$.

When a probability function or density function $f_0$ is not fully
specified, that is, $f_0$ depends some unknown parameters, say
$\boldsymbol{\theta}$, the probabilities $p_1, \cdots, p_k$ depend
on $\boldsymbol{\theta}$.  We can replace $\boldsymbol{\theta}$ with
some estimators such as the maximum likelihood estimator, then
$\mathcal{X}^2_n$ still converges in distribution to a chi-squared
distribution with $k-r-1$ degrees of freedom where $r$ is the
dimension of $\boldsymbol{\theta}$. For more topics and their
developments related to Pearson's test statistics, we refer to
Voinov et al.~\cite{Bala2013}.

When the sample size $n$ is small or $k$ is relatively large, some
expected frequencies $e_i$ may become too small. A variety of
estimates of the discrete probability distribution of Pearson's test
statistics have been discussed in the literature, see, e.g.
Cochran~\cite{Cochran1952}, Yarnold~\cite{Yarnold1970},
Larntz~\cite{Larntz1978}, Lawal~\cite{Lawal980},
Hutchinson~\cite{hutchinson} and references therein. Baglivo et
al.~\cite{Baglivo1992} derived formulas for the exact distributions
and significance levels of Pearson's goodness of fit test
statistics. Cressie and Read~\cite{Cressie1989} provided a
comprehensive review for Pearson's goodness-of-fit test and the
likelihood ratio test.

In this paper, we are interested in the goodness-of-fit test when
both $n$ and $k$ go to infinity, that is, we  allow that $k=k_n$
changes with $n$ and $k_n$ can be even much larger than $n$. We note
that asymptotic normality of Pearson's chi-squared test statistics
has been obtained by Tumanyan~\cite{tumanyan1956} and
Holst~\cite{Holst1972} when $n/k_n\to a\in (0,\infty)$ and some
restrictive conditions are held. A recent work by Rempa{\l}a and
Weso{\l}owski~\cite{Rempala2016} extended this scope by imposing
conditions on the following decomposition of Pearson's test
statistics:
\begin{equation}\label{SS}
\mathcal{X}^2_n=S_{n1}+S_{n2}, ~\mbox{ where }
S_{n1}=\sum^{k_n}_{i=1}\frac{(o_i-e_i)^2}{e_i}-
\sum^{k_n}_{i=1}\frac{o_i-e_i}{e_i},~S_{n2}=\sum^{k_n}_{i=1}\frac{o_i-e_i}{e_i}.
\end{equation}
By assuming that $S_{n2}$ is negligible,  Rempa{\l}a and
Weso{\l}owski~\cite{Rempala2016} showed that $\mathcal{X}^2_n$ is
asymptotically normal if $n^2/k_n\to \infty$ as $n\to\infty$. The
conditions imposed in Rempa{\l}a and
Weso{\l}owski~\cite{Rempala2016} will be discussed further in
Section~\ref{main}. Since the negligibility condition is trivially
true for equiprobable cells, that is, $p_1=\cdots=p_{k_n}$,
$\mathcal{X}^2_n$ has a normal limit, and furthermore, Rempa{\l}a
and Weso{\l}owski~\cite{Rempala2016} showed in this case that
$\mathcal{X}^2_n$, after properly normalized, converges in
distribution to a Poisson distribution if $n^2/k_n\to \lambda\in
(0,\infty)$.

Pearson's chi-squared test has been proven to be unbiased if one
uses equiprobable cells, see, e.g.  Mann and Wald~\cite{Mann1942}
and Cohen and Sackrowitz~\cite{Cohen1975}.  Koehler and
Larntz~\cite{Koehler1980} provided empirical evidence for the
accuracy of the normal approximation when $n^2/k_n$ is reasonably
large.

If one does not use equiprobable cells, Haberman~\cite{Haberman1988}
noted that Pearson's test can be biased when some expected
frequencies become too small. And this is the case if $k_n$ is too
large compared with $n$.  To overcome this drawback,
Zelterman~\cite{zelterman1986, zelterman1987} proposed to use $D^2$
statistic for the test, i.e.  $S_{n1}$ in the decomposition
\eqref{SS}.  Kim et al.~\cite{Kim2009} compared some asymptotic
properties of $\mathcal{X}^2_n$ statistic and $D^2$ statistic for
large sparse multinomial distributions.

In this paper, we investigate the limiting distribution for
Pearson's goodness-of-fit test statistic $\mathcal{X}^2_n$ and $D^2$
statistic. By using  the decomposition \eqref{SS} for Pearson's
goodness-of-fit test statistic we propose some new test statistics
which are more powerful in general.

The rest of the paper is organized as follows. In
Section~\ref{main}, we investigate the limiting distributions of
Pearson's goodness-of-fit test statistics and new test statistics.
In Section~\ref{sim}, we carry out a simulation study to compare the
performance of these test statistics in terms of the size and the
power of the tests.  In Section~\ref{app}, we apply Pearson's
goodness-of-fit test statistics to test whether the winning numbers
from Minnesota Lottery Game Daily 3 were randomly selected with
equal probabilities. Then we summarize the paper with some
concluding remarks. All proofs are given in the {\color{blue}
Supplement}.

\section{Main results}\label{main}

Throughout, we always assume that $k=k_n\to\infty$ as $n\to\infty$.
We adopt some notations as follows.
 The symbol $\td$ denotes the convergence in distribution, $N(0,1)$ denotes a standard normal random variable,
and $\Phi(x)=(2\pi)^{-1/2}\int^x_{-\infty}e^{-t^2/2}dt$ is the
cumulative distribution function of the standard normal. We also
define
\begin{equation}\label{sigma's}
\sigma^2_{n1}=\frac{2(k_n-1)(n-1)}{n}, ~~\sigma_{n2}^2=\frac{1}{n}\big(\sum^{k_n}_{i=1}\frac{1}{p_i}-k_n^2\big), ~~~
\sigma_{n}^2=\sigma_{n1}^2+\sigma_{n2}^2,
\end{equation}
where $\sigma_n^2$, $\sigma_{n1}^2$ and $\sigma_{n2}^2$ are the
variances of $\mathcal{X}^2_n$, $S_{n1}$ and $S_{n2}$, respectively.
In Read and Cressie~\cite{RC1988}, the first three asymptotic
moments have been derived for the so-called power-divergence
statistics which include Pearson's $\mathcal{X}^2_n$ statistic as a
special case. }


We need to impose the following conditions in deriving the limiting distributions for Pearson's test statistic
$\mathcal{X}^2_n$ and some new test statistics that we will propose in the paper:


We first investigate the asymptotic properties of $S_{n1}$  (i.e.
$D^2$ statistic) and $\mathcal{X}_n^2$.
\begin{equation}\label{C3}
\displaystyle\frac{1}{n^2k_n^2}\sum^{k_n}_{i=1}\frac{1}{p_i^2}\to
0\mbox{ as }n\to\infty,
\end{equation}

\begin{equation}\label{C4}
\displaystyle\min\Big(\frac{\sum^{k_n}_{i=1}\frac{1}{p_i^3}-k_n^4}{n^3
\sigma_{n2}^4}I(\sigma_{n2}^2>0), \frac{\sigma_{n2}^2}{k_n}\Big)\to
0 \mbox{ as }n\to\infty.
\end{equation}
\begin{theorem}\label{thm1} If \eqref{C3} holds, then we have
\begin{equation}\label{CLT2}
\frac{S_{n1}-(k_n-1)}{\sigma_{n1}}\td N(0,1)~~~\mbox{ as }
n\to\infty,
\end{equation}
where $\sigma_{n1}$ is defined in \eqref{sigma's}.
\end{theorem}

\begin{theorem}\label{thm2} Under conditions \eqref{C3} and \eqref{C4} we have
\begin{equation}\label{CLT}
\frac{\mathcal{X}^2_n-(k_n-1)}{\sigma_{n}}\td N(0,1)~~~\mbox{ as } n\to\infty,
\end{equation}
where $\sigma_{n}$ is defined in \eqref{sigma's}.
\end{theorem}

Based on the normal approximation \eqref{CLT}, a test with
approximate size $\alpha$ rejects the null hypothesis if
$\mathcal{X}^2_n>k_n-1+\sigma_{n}z_{\alpha}$, where $z_{\alpha}$
denotes the $\alpha$-level critical value of the standard normal
distribution for each $\alpha\in (0,1)$.  Based on \eqref{CLT2}, a
test with approximate size $\alpha$ rejects the null hypothesis if
$S_{n1}>k_n-1+\sigma_{n1}z_{\alpha}$.

A test of size $\alpha$ is said to be unbiased if the power of the test is at least $\alpha$ under alternative hypotheses.
 The test based on the statistic $\mathcal{X}^2_n$ is not unbiased under some alternatives as pointed out by Haberman~\cite{Haberman1988}.
More seriously, our simulation study indicates that the test has a
nearly zero power under some alternatives, that is, the test loses
its power completely in those cases; see Table~\ref{table2}. In
order to understand why this happens, we will look at the
decomposition \eqref{SS} for the goodness-of-fit test statistic
$\mathcal{X}^2_n$.
From Lemma~\ref{lem1} in the {\color{blue} Supplement}, we have
under the null hypothesis $P(E_i)=p_i$ for $1\le i\le k_n$ that
\[
\E (S_{n1}-(k_n-1))=0\mbox{ and }\E(S_{n2})=0,
\]
and under the alternative $H_1$:  $P(E_i)=p_i'$ for $1\le i\le k_n$ that
\begin{equation}\label{bias}
\E\big(S_{n1}-(k_n-1)|H_1\big)=(n-1)\sum^{k_n}_{i=1}\frac{(p_i'-p_i)^2}{p_i}~\mbox{
and }~\E(S_{n2}|H_1)=\sum^{k_n}_{i=1}\frac{p_i'-p_i}{p_i}.
\end{equation}
The variances under $H_1$ can be calculated for both $S_{n1}$ and
$S_{n2}$. Since the rejection region of the goodness-of-fit test is
one-sided,  the test gains its power from a shift to right in
location of the test statistic $\mathcal{X}^2_n$ under the
alternative.  In $S_{n1}$, the effect of a shift is always positive,
but the sign of $\sum^{k_n}_{i=1}\frac{p_i'-p_i}{p_i}$, the location
shift in $S_{n2}$, can be negative. If this shift in location to
left in $S_{n2}$ is overwhelming, the observed values for
$\mathcal{X}^2_n$ can be very small and will result in rejecting the
alternative hypotheses.

Since we are considering the situation when both $n$ and $k_n$ are
large, from \eqref{bias},  $|\E(S_{n2}|H_1)|$ can be very large
compared with $\E\big(S_{n1}-(k_n-1)|H_1\big)$ when $k_n$ is much
larger than $n$.  This indicates that using $|S_{n2}|$ in the test
statistics can be more powerful than $S_{n2}$.  We propose a class
of test statistics $S_{n1}+c|S_{n2}|$, where $c\ge 0$ is a constant.
Their limiting distributions are given as follows.

\begin{theorem}\label{thm3} Under conditions \eqref{C3} and \eqref{C4}  we have as $n\to\infty$
\begin{equation}\label{CLT4}
\sup_{x}\Big|P\Big(\frac{S_{n1}+c|S_{n2}|-(k_n-1)}{\sigma_{n1}}\le
x\Big)-P\Big(Z_1+\frac{c\sigma_{n2}}{\sigma_{n1}}|Z_2|\le
x\Big)\Big|\to 0,
\end{equation}
where $Z_1$ and $Z_2$ are independent random variables with the
standard normal distribution, and $c\ge 0$ is any given constant.
\end{theorem}



For each $s$, define $\Psi(x,s)$ as the cumulative distribution
function of $Z_1+s|Z_2|$, i.e.
\begin{equation}\label{Psi}
\Psi(x,s)=P(Z_1+s|Z_2|\le x)=\sqrt{\frac{2}{\pi}}\int^\infty_0\Phi(x-st)\exp(-t^2/2)dt.
\end{equation}
For each $\alpha\in (0,1)$, let $\psi_{\alpha}(s)$ denote an
$\alpha$-level critical value of $\Psi(\cdot,s)$, that is,
$1-\Psi(\psi_\alpha(s), s)=\alpha$.  The integral in \eqref{Psi} has
no close form solution but it can be evaluated numerically by using
function `integrate' in \textbf{R}. Critical values
$\psi_{\alpha}(s)$ can be solved via the Newton-Raphson method. Note
that $\Psi(x,0)=\Phi(x)$ and thus $\psi_{\alpha}(0)=z_{\alpha}$ for
$\alpha\in (0,1)$.

Three test statistics,  $S_{n1}$,  $\mathcal{X}_n^2$, and
$S_{n1}+c|S_{n2}|$ with $c\ge 0$, can be used to test the null
hypothesis that $P(E_i)=p_i$ for $1\le i\le k_n$, and their
rejection regions at level $\alpha$, according to equations
\eqref{CLT2}, \eqref{CLT} and \eqref{CLT4},  are given by
\[
\mathcal{R}_0=\Big\{S_{n1}>k_n-1+\sigma_{n1}z_{\alpha}\Big\},
\]
\begin{equation}\label{R}
\mathcal{R}=\Big\{\mathcal{X}_n^2>k_n-1+\sigma_nz_{\alpha}\Big\},
\end{equation}
and
 \begin{equation}\label{Rc}
\mathcal{R}_c=\Big\{S_{n1}+c|S_{n2}|>k_n-1+\sigma_{n1}\psi_{\alpha}(\frac{c\sigma_{n2}}{\sigma_{n1}})\Big\}
\end{equation}
for $c\ge 0$. Note that test $\mathcal{R}_0$ can be considered as a
special case of $\mathcal{R}_c$ defined in \eqref{Rc} with $c=0$.

The aforementioned test statistics (or their corresponding rejection
regions) are the same when $p_1=\cdots=p_{k_n}$ since
$\sigma_{n2}^2=0$ and $S_{n2}=0$ in this case.  Note that
Theorem~\ref{thm1} can be considered as a special case of
Theorem~\ref{thm3} with $c=0$, but in Theorem~\ref{thm1} we impose
only condition \eqref{C3} which is less restrictive than conditions
in Theorem~\ref{thm3}.  If we  assume $p_1=\cdots=p_{k_n}$,
condition \eqref{C3} is equivalent to $\lim_{n\to\infty}k_n/n^2=0$.
Immediately we have the following corollary.

\begin{corollary}\label{cor1} Assume that $\{k_n\}$ is a sequence of positive integers such that $k_n\to\infty$
and $k_n=o(n^2)$ as $n\to\infty$. Then under the assumption that
$p_1=\cdots=p_{k_n}$, we have
\[
\frac{\mathcal{X}^2_n-(k_n-1)}{\sigma_{n1}}\td N(0,1)~\mbox{ as }~ n\to\infty.
\]
\end{corollary}

Under the assumption of the equiprobable cells with
$p_1=\cdots=p_{k_n}$, we have
$S_{n2}=0=\sum^{k_n}_{i=1}\frac{o_i-e_i}{e_i}$ for any samples,
regardless of how large for any single term $\frac{o_i-e_i}{e_i}$.
As a remedy, we can assign a weight for each term such that the
weighted sum is not degenerate.  Now we introduce a weighted version
for $S_{n2}$ as follows
\begin{equation}\label{S2bar}
\overline{S}_{n2}=\sum^{k_n}_{i=1}\frac{c_i(o_i-e_i)}{e_i},
\end{equation}
where $c_i\ge 0$ for $1\le i\le k_n$ and $\sum^{k_n}_{i=1}c_i=k_n$.
Obviously, $S_{n2}$ is a special case of $\overline{S}_{n2}$ with
$c_1=\cdots=c_{k_n}=1$.

 We can verify that
\begin{equation}\label{sigma2bar}
\overline\sigma_{n2}^2=\E(\overline{S}_{n2}^2)=\frac{1}{n}\Big(\sum^{k_n}_{i=1}\frac{c_i^2}{p_i}-k_n^2\Big)
~\mbox{ and }~ \E(\overline{S}_{n2})=0;
\end{equation}
see \eqref{varofS2} in the {\color{blue} Supplement}.

Now we propose a new class of test statistics
$S_{n1}+c|\overline{S}_{n2}|$, where $c\ge 0$ is a constant. We need
the following condition for the asymptotic normality for those test
statistics.

\begin{equation}\label{C44}
\displaystyle\min\Big(\frac{\sum^{k_n}_{i=1}\frac{c_i^4}{p_i^3}-k_n^4}{n^3
\overline\sigma_{n2}^4}I(\overline\sigma_{n2}^2>0),
\frac{\overline\sigma_{n2}^2}{k_n}\Big)\to 0 \mbox{ as }n\to\infty.
\end{equation}

\begin{theorem}\label{thm4} Under conditions \eqref{C3} and \eqref{C44}  we have as $n\to\infty$
\begin{equation}\label{CLT5}
\sup_{x}\Big|P\Big(\frac{S_{n1}+c|\overline{S}_{n2}|-(k_n-1)}{\sigma_{n1}}\le
x\Big)-P\Big(Z_1+\frac{c\overline\sigma_{n2}}{\sigma_{n1}}|Z_2|\le
x\Big)\Big|\to 0,
\end{equation}
where $Z_1$ and $Z_2$ are independent random variables with the
standard normal distribution, and $c\ge 0$ is any given constant.
\end{theorem}

For given weights $c_1, \cdots, c_{k_n}$ and constant $c\ge 0$, a
test of size $\alpha$ based on approximation \eqref{CLT5} for test
statistic $S_{n1}+c|\overline{S}_{n2}|$ has the following rejection
region
\begin{equation}\label{Rcbar}
\overline{\mathcal{R}}_c=\{S_{n1}+c|\overline{S}_{n2}|>k_n-1+\sigma_{n1}\psi_{\alpha}(\frac{c\overline\sigma_{n2}}{\sigma_{n1}})\},
\end{equation}
where $\overline\sigma_{n2}$ is defined in \eqref{sigma2bar}.

Of particular interest, we offer a discussion for the selection on
weights $c_1, \cdots, c_{k_n}$ so that $\overline{S}_{n2}$ is
non-degenerate and condition \eqref{C44} is satisfied for the
equiprobable cells. When $p_1=\cdots=p_{k_n}=\frac{1}{k_n}$, we
select weights $c_1, \cdots, c_{k_n}$ such that they are not
identically equal to $1$. This ensures $\overline\sigma_{n2}^2>0$.

A very simple way is to select an integer $k_0$ such that $k_0\sim
hk_n$ for some $h\in (0,1)$ and assign a value $k_n/k_0$ to $k_0$ of
$c_i$'s and $0$ to the remaining $k_n-k_0$ weights. Then for any
integer $r\ge 1$, we have
\[
\sum^{k_n}_{i=1}\frac{c_i^{r+1}}{p_i^r}-k_n^{r+1}=k_n^rk_0\big(\frac{k_n}{k_0}\big)^{r+1}-k_n^{r+1}=
\Big(\big(\frac{k_n}{k_0}\big)^{r}-1\Big)k_n^{r+1}\sim
\big(h^{-r}-1\big)k_n^{r+1}.
\]
Then the first term in the parentheses in \eqref{C44} approximately
equals
\[
\frac{(h^{-3}-1)k_n^{4}}{n\big((h^{-1}-1)k_n^2\big)^2)}\sim
\frac{h^{-3}-1}{(h^{-1}-1)^2}\frac{1}{n}\to 0
\]
as $n\to\infty$. That is, \eqref{C44} holds.



\section{A simulation study}\label{sim}

In this section, we compare the performance of the test statistics
defined in Section~\ref{main} through some simulations.


We first compare the sizes and powers of the test statistics
$\mathcal{X}_n^2$, $S_{n1}$ and $S_{n1}+c|S_{n2}|$ under some
general null hypotheses.  Then we compare the performance of
$\mathcal{X}_n^2$ and $S_{n1}+c|\overline{S}_{n2}|$ under
equiprobable cells with $p_1=\cdots=p_{k_n}$.

Firstly,  we consider the following five tests, including
$\mathcal{R}$, $\mathcal{R}_0$, $\mathcal{R}_1$, $\mathcal{R}_3$ and
$\mathcal{R}_5$ as defined in \eqref{R} and \eqref{Rc} with
selection of $\alpha=0.05$ and several combinations of $n$ and
$k_n$. For each case, the simulation is repeated $10000$ times by
using \textbf{R} package, and the sizes and powers of these tests
are estimated.


We assume $k_n$ is an even integer and define
\begin{equation}\label{distribution1}
p_1=\cdots=p_{\frac{k_n}{2}}=\frac{r}{k_n},~~ p_{\frac{k_n}{2}+1}=\cdots=p_{k_n}=\frac{2-r}{k_n}
\end{equation}
for $r\in (0,2)$. For given $n$ and $k_n$,  each of the above
probability distributions is uniquely  determined by $r$. We note
that $p_1=\cdots=p_{k_n}=\frac1{k_n}$ if and only if $r=1$.

Table~\ref{table1} contains estimated sizes for the five tests
$\mathcal{R}$, $\mathcal{R}_0$, $\mathcal{R}_1$, $\mathcal{R}_3$ and
$\mathcal{R}_5$ with $n=100$, $1000$ and some selected values for
$k_n$. For each combination of ($n$, $k_n)$, we take three
probability distributions from \eqref{distribution1} with $r=0.1$,
$0.2$, $0.6$ and $1.0$, respectively. When $r=1.0$, all the five
tests are the same. From Table~\ref{table1} the estimated sizes for
all five tests are very close to the nominal level $0.05$, and thus,
we conclude all the five tests perform very well in terms of the
accuracy in type I error.

\begin{figure}
\centering
\includegraphics[height=.2\textheight, width=1\textwidth]{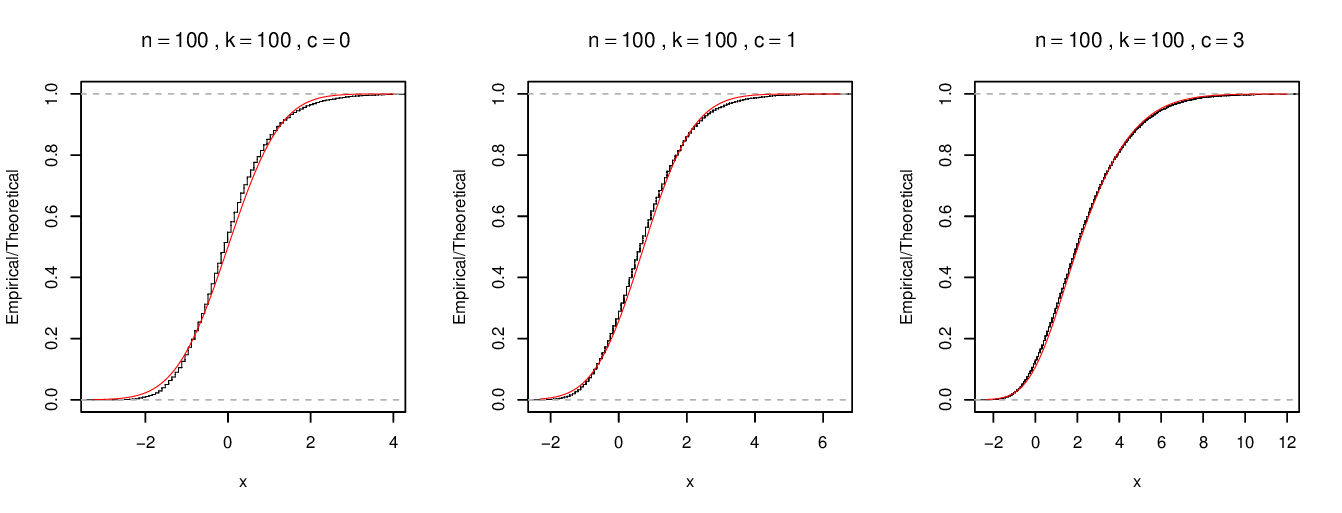}

\includegraphics[height=.2\textheight, width=1\textwidth]{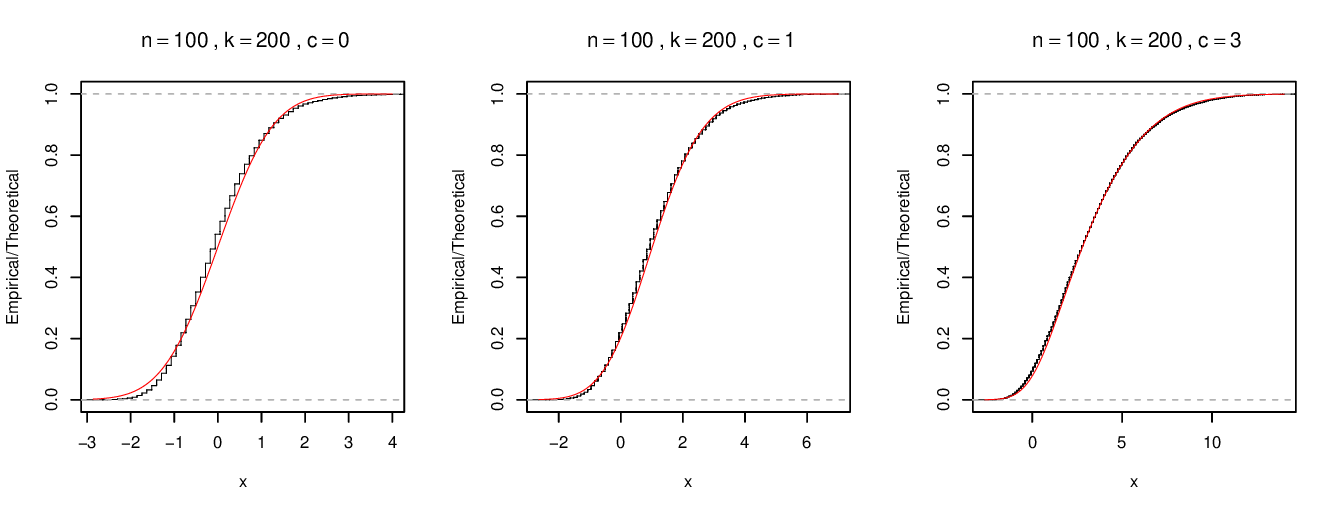}

\includegraphics[height=.2\textheight, width=1\textwidth]{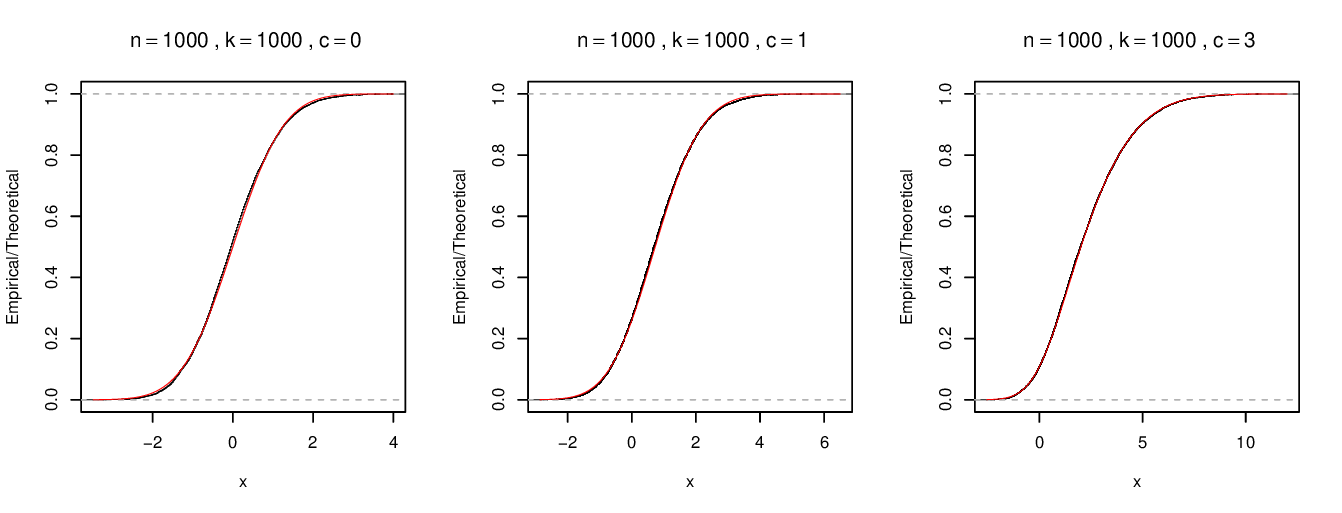}

\includegraphics[height=.2\textheight, width=1\textwidth]{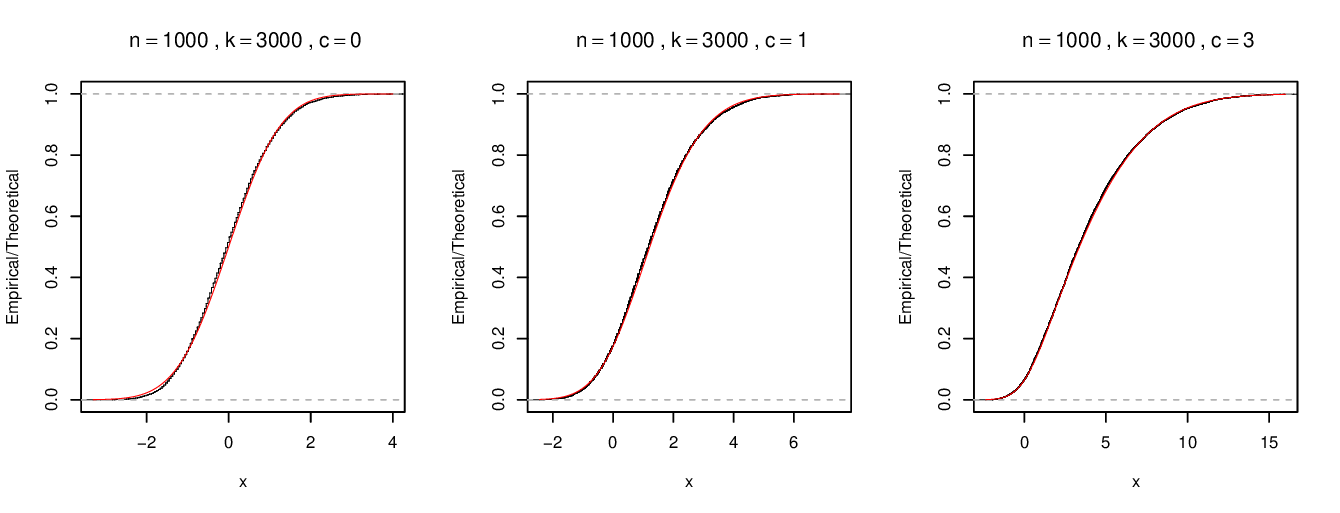}
\caption {Plots of the empirical distribution for the standardized
test statistics $(S_{n1}+c|S_{n2}|-(k_n-1))/\sigma_{n1}$ and their
theoretical limiting cumulative distribution $\Psi(x,
c\sigma_{n2}/\sigma_{n1})$ under the null hypothesis from family
\eqref{distribution1} with $r=0.2$. $\Psi$ is defined in
\eqref{Psi}. We select $c=0, 1,$ and $3$, corresponding to the tests
$\mathcal{R}_0$, $\mathcal{R}_1$ and $\mathcal{R}_3$. In these
plots, the red/smooth lines represent the theoretical cumulative
distributions $\Psi(x, c\sigma_{n2}/\sigma_{n1})$. }
 \label{figure1}
\end{figure}

To assess the overall performance of the distributional
approximations to the standardized test statistics
$(S_{n1}+c|S_{n2}|-(k_n-1))/\sigma_{n1}$  under the null hypothesis,
we compare the empirical distributions of the test statistics based
on 10000 samples and their theoretical cumulative distribution
functions under the null hypothesis with a distribution from family
\eqref{distribution1}.  Figure~\ref{figure1} contains plots for both
the empirical distributions and  the approximate distributions of
the test statistics with $c=0$, $1$, $3$. The parameter of the
distribution under $H_0$ is set to be $r=0.2$.  We discover from
Figure~\ref{figure1} that all three theoretical distributions for
the test statistics fit the empirical distributions very well,  and
their accuracies improve when sample size is getting large. Results
for other distributions are similar and are not reported here.

To estimate the power for these tests, for each combination of $n$
and $k_n$, we choose probability distribution
 \eqref{distribution1} as the null hypothesis with $r=0.2$ (or $r=0.6$), and use probability distribution
 \eqref{distribution1} with $r=0.2\pm 0.1$ (or accordingly, $r=0.6\pm 0.1$)
  as alternative hypotheses from which random samples are generated.  Table~\ref{table2} lists the estimated powers for the five tests.
  Surprisingly, the power of test $\mathcal{R}$
is nearly zero when the value of $r$ in the true alternative
hypothesis is smaller than the value of $r$ specified in the null
hypothesis, that is,  Pearson's goodness-of-fit test is seriously
biased in those cases. As we have explained below equation
\eqref{SS}, this is mainly due to
$\sum^{k_n}_{i=1}\frac{p_i'-p_i}{p_i}$.
 For example, if $r=0.2$ for the distribution under the null hypothesis and $r=0.1$ under the alternative, we have
 \[
 \sum^{k_n}_{i=1}\frac{p_i'-p_i}{p_i}=(\frac{-0.1}{0.2}+\frac{0.1}{1.8})\frac{k_n}{2}=-\frac{2k_n}{9}.
 \]
We can also estimate the standard deviation of $\mathcal{X}^2_n$
under the alternative hypothesis and find out that it is much
smaller than order $k_n$. This explains the incapability of
Pearson's goodness-of-fit test in detecting an alternative in this
case. We also notice that the performance of $\mathcal{R}$ is quite
regular when  $r=0.2$ for the distribution under the null hypothesis
and $r=0.3$ under the alternative. Since
$\sum^{k_n}_{i=1}\frac{p_i'-p_i}{p_i}=\frac{2k_n}{9}>0$ in this
case,  Pearson's goodness-of-fit test gains its power. In both
examples, the power of test $\mathcal{R}_c$ increases with $c$. In
the first example,  test $\mathcal{R}_c$ is superior to
$\mathcal{R}$ for all $c\ge 0$.  In the second example,  test
$\mathcal{R}_c$ outperforms $\mathcal{R}$ when $c\ge 3$.

It seems plausible that test $\mathcal{R}_c$ with $c>0$ improves
upon $\mathcal{R}$ when $\sum^{k_n}_{i=1}\frac{p_i'-p_i}{p_i}$ is
quite different from zero. It is interesting to know how much
improvement can be made when $\sum^{k_n}_{i=1}\frac{p_i'-p_i}{p_i}$
is zero or very close to zero. To make an empirical comparison, we
introduce a new family of probability distributions. For
convenience, we assume that $k_n$ is divisible by $4$ and define a
family of probability distributions
\begin{equation}\label{distribution2}
p_1'=\cdots=p_{\frac{k_n}{4}}'=\frac{1.5r'}{k_n},
~p_{\frac{k_n}{4}+1}'=\cdots=p_{\frac{k_n}{2}}'=\frac{0.5r'}{k_n},~
p_{\frac{k_n}{2}+1}'=\cdots=p_{k_n}'=\frac{2-r'}{k_n}
\end{equation}
for $r'\in (0,2)$. For given $n$ and $k_n$, each probability
distribution above is uniquely determined by $r'$. It is easy to see
that for a probability distribution $(p_1, \cdots, p_{k_n})$ from
\eqref{distribution1} and a probability distribution
 $(p_1', \cdots,
p_{k_n}')$ from \eqref{distribution2},
$\sum^{k_n}_{i=1}\frac{p_i'-p_i}{p_i}=0$ if $r=r'$.
Table~\ref{table3} includes estimated powers of the five tests for
several combinations of $n$ and  $k_n$ with $r=r'=0.6$, $1.4$. The
probability distribution under the null hypothesis is from family
\eqref{distribution1} with parameter $r$ and  the true probability
distribution under the alternative is from \eqref{distribution2}
with parameter $r'=r$. From Table~\ref{table3}, the power of
$\mathcal{R}_c$ decreases with $c$ for large $c$. We note that the
constant $c$ represents the weight of $|S_{n2}|$ we take into
account in the test,  and $S_{n1}$ has always a positive shift in
location under the alternative, and thus it is more likely to detect
the alternative if the weight of $|S_{n2}|$ in the test is smaller.
In other words, increasing the weight $c$ can decrease the power of
$\mathcal{R}_c$ in this case. Therefore, we do not recommend to use
a large $c$ in general.   In Table~\ref{table3}, we have used least
favorable distributions to the use of $S_{n2}$ under alternatives
since the expectations of $S_{n2}$ under the alternatives are zero.
Overall, the performance of test $\mathcal{R}_0$ is slightly better
than test $\mathcal{R}_1$ from Table~\ref{table3}.

Now we compare $\mathcal{X}_n^2$ and $S_{n1}+c|\overline{S}_{n2}|$
for the case of equiprobable cells.  Recall that
$\mathcal{X}_n=S_{n1}$ in this case. We define $\overline{S}_{n2}$
by using the method discussed at the end of Section~\ref{main}, that
is, we define $k_0=0.8k_n$ and set
$c_i=\frac{k_n}{k_0}=\frac{1}{0.8}=1.25$ for $1\le i\le k_0$ and $0$
otherwise.  This time, we consider only two tests, $\mathcal{R}_0$
and $\overline{\mathcal{R}}_1$, as defined in \eqref{R} and
\eqref{Rcbar} with $c=1$. For several combinations of $n$ and $k_n$,
the sizes for the two tests are estimated based on $10000$
replicates. The powers for the two tests are also estimated when the
distributions under the alternatives are from family
\eqref{distribution1} with $r=0.8$, $1.2$ and $1.4$ or from family
\eqref{distribution2} with $r'=0.8$, $1.1$ and $1.2$.  The estimated
sizes and powers are reported in Table~\ref{table4}.

From Table~\ref{table4}, we conclude that the sizes for both
$\mathcal{R}_1$ and $\overline{\mathcal{R}}_1$ are close to the
nominal level $0.05$, and in general, $\overline{\mathcal{R}}_1$ is
more powerful than $\mathcal{R}_1$. These empirical results are
consistent with Theorem~\ref{thm4}, and adding the term
$\overline{S}_{n2}$ defined in \eqref{S2bar}, a non-trivial linear
combination of the terms $(o_i-e_i)/e_i$, can improve the power of
the test significantly.

Next, we extend our comparison of $S_{n1}$ and
$S_{n1}+c|\overline{S}_{n2}|$ to some none-equiprobable cases. We
also use the method discussed at the end of Section~\ref{main} to
define $\overline{S}_{n2}$ by setting $k_0=0.40k_n$ this time. We
consider the distributions from family \eqref{distribution2} and use
the same settings as in Table~\ref{table3}.  Only for an
illustration purpose,  we demonstrate the weighted test statistics
can improve the power of the test $\mathcal{R}_1$ when the
expectation of $S_{n2}$ under $H_1$ in \eqref{bias} is zero.   Both
the sizes and powers for tests $\mathcal{R}_0$ and $\mathcal{R}_1$
are reported in Table~\ref{table5}. From the table, we see that both
the tests maintain reasonable sizes for all combinations of $n$ and
$k_n$.  From Tables~\ref{table3} and \ref{table5}, we can conclude
that test $\overline{\mathcal{R}}_1$ performs significantly better
than $\mathcal{R}_0$ and $\mathcal{R}_1$ in terms of power.

Finally, we compare the performance of these tests under sparsity.
To this end, we introduce a class of distributions as follows
\begin{equation}\label{distribution3}
p_1=\cdots=p_{0.95k_n}=\frac{r}{8k_n},~~
p_{0.95k_n+1}=\cdots=p_{k_n}=\frac{160-19r}{8k_n}
\end{equation}
for $r\in (0,8)$, where $k_n$ is a multiple of $20$. We see that
only $5\%$ of probabilities $p_i$'s take a larger value
$\frac{160-19r}{8k_n}$ in \eqref{distribution3}. In our study, we
select $r=2$ for the distribution under the null hypothesis and
estimate the sizes of the five tests considered in
Table~\ref{table1} and estimate the powers of the tests under the
alternatives $r=1$ and $r=3$ for some combinations of $n$ and $k_n$.
The estimated sizes and powers are reported in Table~\ref{table6}.

Results in Table~\ref{table6} are quite similar to
Tables~\ref{table1} and \ref{table2} in the following aspects:
\textbf{a.} the estimated type I errors are close to nominal level
$0.05$ for all five tests;  \textbf{b.} Pearson's test $\mathcal{R}$
loses its power totally for some distributions under the alternative
while  tests $\mathcal{R}_1$, $\mathcal{R}_3$, and $\mathcal{R}_5$
outperform with large powers. Although test $\mathcal{R}_0$ is
better than Pearson's test, its overall performance is not quite
satisfactory. For example, for some distributions under the
alternative,  its powers are smaller than the type I errors. We
examine the results in Table~\ref{table6} when $n=100$ with $r=1$
under the alternative and find out that  the power of the test
decreases from $0.0416$ to $0.0246$ when $k_n$ increases from $100$
to $400$. The same phenomenon can also be observed when $n=1000$.

 It is worth mentioning that conditions \eqref{C3} and \eqref{C4}
that ensure the asymptotic normality of $S_{n1}$ and
$\mathcal{X}^2_n$ may be moderately violated if $k_n$ is too large
compared with $n$. This is the case for some combinations of $n$ and
$k_n$ and for some distributions used in Table~\ref{table6}. In our
study, the sizes (type I errors) of all five tests are reasonably
close to the nominal level $0.05$; see Tables~\ref{table1} and
\ref{table6}.   In terms of power, test $\mathcal{R}_1$,
$\mathcal{R}_3$, and $\mathcal{R}_5$ are also very robust as they
gain good powers from Tables~\ref{table2} and \ref{table6}.

To conclude this section, we present more discussion on selection of
$c$. Our simulation study indicates that there is no answer for
optimal section of $c$ in general.  As we have pointed out, $c$
represents the weight of $|S_{n2}|$ in test $\mathcal{R}_c$.  When
$c$ is large, the test $\mathcal{R}_c$ is almost the same as
$\{|S_{n2}|/\sigma_{n2}>z_{\alpha/2}\}$, where $z_{\alpha/2}$ is the
$\alpha/2$-level critical value of the standard normal distribution.
On the one hand, the power of test $\mathcal{R}_c$ increases with
$c$ in most cases in Tables~\ref{table2} and \ref{table6}. By
comparing the powers of the test $\mathcal{R}_c$ for various values
of $c$ in our simulation study, we find out that the increment in
the power for the test $\mathcal{R}_c$ is very limited when $c$ is
larger than $3$, and the power of $\mathcal{R}_2$ is close to that
of $\mathcal{R}_3$ in most cases. On the other hand,
Table~\ref{table3} indicates that one may prefer to employ a test
$\mathcal{R}_c$ with a smaller value $c$ in the worst scenarios such
as those distributions given in \eqref{distribution2}. We observe
from Table~\ref{table3} that the power of $\mathcal{R}_1$ is very
close to that of $\mathcal{R}_0$ in most cases. Our simulation study
also shows that the power of $\mathcal{R}_2$ is only slightly
smaller than that of $\mathcal{R}_1$ in most cases. Intuitively, the
power of test $\mathcal{R}_c$ depends on the probability
distributions under both the null and alternative hypotheses as well
as the relative convergence rate of $n$ and $k_n$. A theoretical
investigation on how the power function of the test $\mathcal{R}_c$
depends on these factors can be very helpful but may be very
complicated. In practice, the distributions under the alternative
are unknown, and the optimal choice of $c$ that works for all
distributions does not exist. To balance different situations, one
can use tests $\mathcal{R}_1$ or $\mathcal{R}_2$. As a general
recommendation, one can calculate tests $\mathcal{R}$,
$\mathcal{R}_0$, $\mathcal{R}_1$ and $\mathcal{R}_2$ for comparison
purpose. One should be cautious about accepting the null hypothesis
based on $\mathcal{R}$ or $\mathcal{R}_0$ since the powers of the
two tests may be much smaller than their sizes or type I errors. In
other words, tests $\mathcal{R}$ and $\mathcal{R}_0$ may reject the
alternative hypotheses with a probability close to one when the null
hypotheses are not true.


\null\vspace{-50pt}


\begin{table}
{
\footnotesize
\caption{\textbf{Estimated Sizes of Tests (Nominal Level $\alpha=0.05$)}: Probability distributions under null hypotheses and the true distributions that are used to generate samples are from family \eqref{distribution1} with different values for parameter $r$}
\label{table1}
\centering
\begin{tabular}{|lr|r|r|r|r|r|r|r|r}
 \hline
  Sample &  &Distribution&True & \multicolumn{5}{c|}{Probability of Rejecting $H_0$  for Tests}\\
     \cline{5-9}
  Size ($n$)& $k_n$& under $H_0$&distribution &~~~~$\mathcal{R}$& ~~~$\mathcal{R}_0$&~~~$\mathcal{R}_1$
  & ~~~$\mathcal{R}_3$&~~~$\mathcal{R}_5$\\
    \hline
 100 & 50 & $r=0.1$ & $r=0.1$ & 0.0679 & 0.0638 & 0.0642 & 0.0536 & 0.0500\\
 100 & 50 & 0.2 & 0.2 & 0.0663 & 0.0614 & 0.0668 & 0.0633 & 0.0599\\
 100 & 50 & 0.6 & 0.6 & 0.0609 & 0.0593 & 0.0628 & 0.0672 & 0.0674\\
 100 & 50 & 1.0 & 1.0 & 0.0576 & 0.0576 & 0.0576 & 0.0576 & 0.0576\\
 \cline{1-9}
  100 & 100 & $r=0.1$ & $r=0.1$ & 0.0610 & 0.0595 & 0.0569 & 0.0482 & 0.0450\\
  100 & 100 & 0.2 & 0.2 & 0.0667 & 0.0610 & 0.0594 & 0.0567 & 0.0526\\
  100 & 100 & 0.6 & 0.6 & 0.0579 & 0.0562 & 0.0594 & 0.0616 & 0.0642\\
  100 & 100 & 1.0 & 1.0 & 0.0531 & 0.0531 & 0.0531 & 0.0531 & 0.0531\\
 \cline{1-9}
 100 & 200 & $r=0.1$ & $r=0.1$ & 0.0644 & 0.0679 & 0.0567 & 0.0491 & 0.0413\\
 100 & 200 & 0.2 & 0.2 & 0.0619 & 0.0579 & 0.0622 & 0.0536 & 0.0544\\
 100 & 200 & 0.6 & 0.6 & 0.0580 & 0.0550 & 0.0606 & 0.0579 & 0.0539\\
 100 & 200 & 1.0 & 1.0 & 0.0691 & 0.0691 & 0.0691 & 0.0691 & 0.0691\\
 \cline{1-9}
 1000 & 300 & $r=0.1$ & $r=0.1$ & 0.0571 & 0.0543 & 0.0574 & 0.0521 &0.0519\\
 1000 & 300 & 0.2 & 0.2 & 0.0552 & 0.0523 & 0.0567 & 0.0562 & 0.0538\\
 1000 & 300 & 0.6 & 0.6 & 0.0515 & 0.0518 & 0.0527 & 0.0536 & 0.0556\\
 1000 & 300 & 1.0 & 1.0 & 0.0514 & 0.0514 & 0.0514 & 0.0514 & 0.0514\\
 \cline{1-9}
 1000 & 1000 & $r=0.1$ & $r=0.1$ & 0.0570 & 0.0594 & 0.0583 & 0.0533 &0.0543 \\
 1000 & 1000 & 0.2 & 0.2 & 0.0525 & 0.0548 & 0.0526 & 0.0523 & 0.0489\\
 1000 & 1000 & 0.6 & 0.6 & 0.0538 & 0.0534 & 0.0526 & 0.0505 & 0.0516\\
 1000 & 1000 & 1.0 & 1.0 & 0.0530 & 0.0530 & 0.0530 & 0.0530 & 0.0530\\
 \cline{1-9}
 1000 & 3000 & $r=0.1$ & $r=0.1$ & 0.0542 & 0.0600 & 0.0515 & 0.0501 &0.0505\\
 1000 & 3000 & 0.2 & 0.2 & 0.0549 & 0.0568 & 0.0570 & 0.0511 & 0.0510\\
 1000 & 3000 & 0.6 & 0.6 & 0.0544 & 0.0546 & 0.0561 & 0.0553 & 0.0532\\
 1000 & 3000 & 1.0 & 1.0 & 0.0565 & 0.0565 & 0.0565 & 0.0565 & 0.0565\\
 \cline{1-9}
 1000 & 10000 & $r=0.1$ & $r=0.1$ & 0.0485 & 0.0648 & 0.0447 & 0.0468 & 0.0472\\
 1000 & 10000 & 0.2 & 0.2 & 0.0544 & 0.0614 & 0.0506 & 0.0493 & 0.0494\\
 1000 & 10000 & 0.6 & 0.6 & 0.0520 & 0.0590 & 0.0567 & 0.0491 & 0.0486\\
 1000 & 10000 & 1.0 & 1.0 & 0.0567 & 0.0567 & 0.0567 & 0.0567 & 0.0567\\
   \hline
\end{tabular}}
\end{table}


\begin{table}
{
\footnotesize
 \caption{\textbf{Estimated Powers
of Tests ($\alpha=0.05$)}: Probability distributions under null
hypotheses and the true distributions that are used to generate
samples are from family \eqref{distribution1} with different values
for parameter $r$}
\label{table2} \centering
\begin{tabular}{|lr|r|r|r|r|r|r|r|r}
 \hline
  Sample &  &Distribution& True & \multicolumn{5}{c|}{Probability of Rejecting $H_0$  for Tests}\\
     \cline{5-9}
  Size ($n$)& $k_n$& under $H_0$&distribution&~~~~$\mathcal{R}$& ~~~$\mathcal{R}_0$&~~~$\mathcal{R}_1$
  & ~~~$\mathcal{R}_3$&~~~$\mathcal{R}_5$\\
    \hline
100 & 50 & $r=0.1$ & $r=0.05$ & 0.0042 & 0.0455 & 0.1044 & 0.1408 & 0.1315\\
100 & 50 & 0.2 & 0.1 & 0.0041 & 0.0677 & 0.1998 & 0.3646 & 0.3864\\
100 & 50 & 0.2 & 0.3 & 0.3207 & 0.1377 & 0.2608 & 0.3679 & 0.3884\\
100 & 50 & 0.6 & 0.5 & 0.0436 & 0.0699 & 0.0909 & 0.1316 & 0.1614\\
100 & 50 & 0.6 & 0.7 & 0.1247 & 0.0856 & 0.1066 & 0.1463 & 0.1768\\
    \cline{1-9}
100 & 200 & $r=0.1$ & $r=0.05$ & 0.0025 & 0.0372 & 0.1120 & 0.1096 & 0.0893\\
100 & 200 & 0.2 & 0.1 & 0.0010 & 0.0413 & 0.2627 & 0.3577 & 0.3730\\
100 & 200 & 0.2 & 0.3 & 0.3890 & 0.1156 & 0.3178 & 0.3831 & 0.3935\\
100 & 200 & 0.6 & 0.5 & 0.0231 & 0.0573 & 0.0924 & 0.1526 & 0.1743\\
100 & 200 & 0.6 & 0.7 & 0.1399 & 0.0723 & 0.1111 & 0.1678 & 0.1946\\
   \cline{1-9}
1000 & 300 & $r=0.1$ & $r=0.05$ & 0.0000 & 0.0936 & 0.7955 & 0.9723 & 0.9823\\
1000 & 300 & 0.2 & 0.1 & 0.0001 & 0.2739 & 0.9465 & 0.9996 & 1.0000\\
1000 & 300 & 0.2 & 0.3 & 0.8806 & 0.3361 & 0.8425 & 0.9876 & 0.9951\\
1000 & 300 & 0.6 & 0.5 & 0.0369 & 0.1202 & 0.2360 & 0.5269 & 0.7207\\
1000 & 300 & 0.6 & 0.7 & 0.2917 & 0.1464 & 0.2591 & 0.5150 & 0.6971\\
   \cline{1-9}
1000 & 1000 & $r=0.1$ & $r=0.05$ & 0.0000 & 0.0577&0.9308 & 0.9825 & 0.9856\\
1000 & 1000 & 0.2 & 0.1 & 0.0000 & 0.1123 & 0.9937 & 0.9999 & 1.0000\\
1000 & 1000 & 0.2 & 0.3 & 0.9678 & 0.2041 & 0.9463 & 0.9953 & 0.9968\\
1000 & 1000 & 0.6 & 0.5 & 0.0072 & 0.0814 & 0.2860 & 0.7165 & 0.8531\\
1000 & 1000 & 0.6 & 0.7 & 0.3585 & 0.1000 & 0.2880 & 0.6896 & 0.8311\\
   \cline{1-9}
1000 & 3000 & $r=0.1$ & $r=0.05$ & 0.0000 & 0.0423 & 0.9741 & 0.9874
& 0.9880\\
1000 & 3000 & 0.2 & 0.1 & 0.0000 & 0.0657 & 0.9996 & 1.0000 & 1.0000\\
1000 & 3000 & 0.2 & 0.3 & 0.9951 & 0.1545 & 0.9902 & 0.9974 & 0.9979\\
1000 & 3000 & 0.6 & 0.5 & 0.0004 & 0.0632 & 0.4394 & 0.8597 & 0.9174\\
1000 & 3000 & 0.6 & 0.7 & 0.5278 & 0.0826 & 0.4314 & 0.8272 & 0.8948\\
   \cline{1-9}
1000 & 10000 & $r=0.1$ & $r=0.05$ & 0.0000 & 0.0431 & 0.9838 & 0.9883 & 0.9891\\
1000 & 10000 & 0.2 & 0.1 & 0.0000 & 0.0476 & 1.0000 & 1.0000 & 1.0000\\
1000 & 10000 & 0.2 & 0.3 & 0.9974 & 0.1292 & 0.9952 & 0.9967 & 0.9969\\
1000 & 10000 & 0.6 & 0.5 & 0.0000 & 0.0575 & 0.7122 & 0.9226 & 0.9377\\
1000 & 10000 & 0.6 & 0.7 & 0.7677 & 0.0777 & 0.6766 & 0.8997 & 0.9165\\
   \hline
\end{tabular}}
\end{table}


\begin{table}
{
\footnotesize
\null\vspace{-60pt} \caption{\textbf{Estimated Powers
of Tests ($\alpha=0.05$)}: Probability distributions under null
hypotheses and the true distributions that are used to generate
samples are from family \eqref{distribution1} with parameter $r$ and
family \eqref{distribution2} with parameter $r'$, respectively}
\label{table3} \centering
\begin{tabular}{|lr|r|r|r|r|r|r|r|r}
 \hline
  Sample &  &Distribution& Distribution& \multicolumn{5}{c|}{Probability of Rejecting $H_0$  for Tests}\\
     \cline{5-9}
  Size ($n$)& $k_n$& under $H_0$&under $H_1$&~~~~$\mathcal{R}$& ~~~$\mathcal{R}_0$&~~~$\mathcal{R}_1$
  & ~~~$\mathcal{R}_3$&~~~$\mathcal{R}_5$\\
\hline
100 & 100 & $r=$0.6 & $r'=$0.6 & 0.1562 & 0.1511 & 0.1536 & 0.1400 & 0.1173\\
100 & 100 & 1.4 & 1.4 & 0.3179 & 0.3418 & 0.3386 & 0.2895 & 0.2168\\
\cline{1-9}
100 & 200 & $r=$0.6 & $r'=$0.6 &0.1197 & 0.1238 & 0.1215 & 0.0992 & 0.0808\\
100 & 200 & 1.4 & 1.4 & 0.1940 & 0.2260 & 0.2241 & 0.1647 & 0.1185\\
\cline{1-9}
1000 & 300 & $r=$0.6 & $r'=$0.6 & 0.8606 & 0.8739 & 0.8716 & 0.8513 & 0.8083\\
1000 & 300 & 1.4 & 1.4 & 0.9999 & 1.0000 & 1.0000 & 1.0000 & 1.0000\\
\cline{1-9}
1000 & 1000 & $r=$0.6 & $r'=$0.6 & 0.4694 & 0.4939 & 0.4833 & 0.4016 & 0.2968\\
1000 & 1000 & 1.4 & 1.4 & 0.9713 & 0.9758 & 0.9721 & 0.9403 & 0.8441\\
\cline{1-9}
1000 & 3000 & $r=$0.6 & $r'=$0.6 & 0.2252 & 0.2575 & 0.2370 & 0.1556 & 0.1104\\
1000 & 3000 & 1.4 & 1.4 & 0.6195 & 0.7019 & 0.6605 & 0.4225 & 0.2482\\
\cline{1-9}
1000 & 10000 & $r=$0.6 & $r'=$0.6 & 0.1156 & 0.1547 & 0.1292 & 0.0794 & 0.0671\\
1000 & 10000 & 1.4 & 1.4 & 0.2212 & 0.3412 & 0.2761 & 0.1275 & 0.0924\\
\hline
\end{tabular}}
\end{table}


\begin{table}
{
\footnotesize
 \caption{\textbf{Estimated Sizes
and Powers for Tests $\mathcal{R}_0$ and $\overline{\mathcal{R}}_1$
for Equiprobable Cells ($\alpha=0.05$)}.  The distributions under
alternatives are from family \eqref{distribution1} with parameter
$r$ and from family \eqref{distribution2} with parameter $r'$,
respectively} \label{table4} \centering
\begin{tabular}{|lr|l|r|r|r|r|r|r|r|r}
 \hline
  Sample &  &    &    & \multicolumn{3}{c|}{Power under family \eqref{distribution1}}& \multicolumn{3}{c|}{Power under family \eqref{distribution2}}\\
          \cline{5-10}
  Size ($n$)& $k_n$&Tests  &Sizes  &$r=0.8$&$r=1.2$&$r=1.4$
  &$r'=0.8$&$r'=1.1$&$r'=1.2$\\
\hline
100 & 100 & $\mathcal{R}_0$               &0.0531 &0.0964 &0.0900&0.3009&0.2013&0.3030&0.4264\\
100 & 100 & $\overline{\mathcal{R}}_1$  &0.0593 &0.1306 &0.1262&0.4449&0.2544&0.3208&0.4656\\
\hline
100 & 200 & $\mathcal{R}_0$               &0.0691 &0.0976 &0.0973&0.2342&0.1830&0.2428&0.3349\\
100 & 200 & $\overline{\mathcal{R}}_1$  &0.0604 &0.1256 &0.1140&0.3720&0.2084&0.2234&0.3450\\
\hline
1000 & 300 & $\mathcal{R}_0$              &0.0514 &0.4804 &0.4739&0.9999&0.9978&0.9998&1.0000\\
1000 & 300 & $\overline{\mathcal{R}}_1$ &0.0556 &0.6341 &0.6264&1.0000&0.9990&0.9998&1.0000\\
\hline
1000 & 1000 & $\mathcal{R}_0$             &0.0530 &0.2249 &0.2238&0.9515&0.9501 &0.9998&0.9945\\
1000 & 1000 & $\overline{\mathcal{R}}_1$&0.0552 &0.5072 &0.4986&0.9993&0.9604 &0.9998&0.9988\\
\hline
1000 & 3000 & $\mathcal{R}_0$             &0.0565 &0.1384 &0.1435&0.6444&0.4710&0.6459&0.8444\\
1000 & 3000 & $\overline{\mathcal{R}}_1$&0.0561 &0.5650 &0.5587&0.9984&0.8342&0.7316&0.9675\\
\hline
1000 & 10000& $\mathcal{R}_0$             &0.0567 &0.0963 &0.0972 &0.3176 &0.2268&0.3136 &0.4599\\
1000 & 10000& $\overline{\mathcal{R}}_1$&0.0494 &0.7127 &0.7153 &1.0000 &0.8199&0.4869 &0.9097\\
\hline
\end{tabular}}
\end{table}


\begin{table}[h!]
{
\footnotesize \null\vspace{-60pt} \caption{\textbf{Estimated Sizes
and Powers of Tests $\mathcal{R}_0$ and $\overline{\mathcal{R}}_1$
($\alpha=0.05$)}: Probability distributions under null hypotheses
are from  family \eqref{distribution1} with parameter $r$ and the
distributions under alternatives are from family
\eqref{distribution2} with parameter $r'$} \label{table5} \centering
\begin{tabular}{|lr|r|r|r|r|r|r|r}
 \hline
  Sample &  &Distribution &\multicolumn{2}{c|}{Sizes for Tests} &Distribution &\multicolumn{2}{c|}{Powers for Tests}\\
     \cline{4-5}\cline{7-8}
  Size ($n$)& $k_n$& under $H_0$ & ~~~$\mathcal{R}_0$&~~~$\overline{\mathcal{R}}_1$ & under $H_1$ &~~$\mathcal{R}_0$ &~~~$\overline{\mathcal{R}}_1$\\
\hline
100 & 100 & $r=$0.6 &  0.0562& 0.0612 &$r'=$0.6 & 0.1511 & 0.1844 \\
100 & 100 & 1.4 & 0.0612 &0.0665 & 1.4 & 0.3418 & 0.4478\\
\cline{1-8}
100 & 200 & $r=$0.6 & 0.0550 & 0.0515 &$r'=$0.6 & 0.1238 & 0.1525\\
100 & 200 & 1.4 & 0.0564 & 0.0606 &1.4 & 0.2260 & 0.3704 \\
\cline{1-8}
1000 & 300 & $r=$0.6 & 0.0518 & 0.0560 &$r'=$0.6 & 0.8739 & 0.9227\\
1000 & 300 & 1.4 &  0.0537 & 0.0548 &1.4 & 1.0000 & 1.0000 \\
\cline{1-8}
1000 & 1000 & $r=$0.6 & 0.0534 & 0.0547 & $r'=$0.6 & 0.4939 & 0.7565\\
1000 & 1000 & 1.4 & 0.0573 & 0.0560 & 1.4 & 0.9758 & 0.9995\\
\cline{1-8}
1000 & 3000 & $r=$0.6 & 0.0546 & 0.0542&$r'=$0.6 & 0.2575 & 0.6527 \\
1000 & 3000 & 1.4 & 0.0515 & 0.0547 & 1.4 & 0.7019 & 0.9949\\
\cline{1-8}
1000 & 10000 & $r=$0.6 &0.0590 & 0.0497  &$r'=$0.6 & 0.1547 & 0.6164\\
1000 & 10000 & 1.4 &0.0576 & 0.0505 & 1.4 & 0.3412 & 0.9947\\
\hline
\end{tabular}}
\end{table}


\begin{table}[h!]
{
\footnotesize
 \null\vspace{-35pt} \caption{\textbf{Estimated Sizes
and Powers of Tests under Sparsity (Nominal Level $\alpha=0.05$)}:
Probability distributions under null hypotheses and the true
distributions that are used to generate samples are from family
\eqref{distribution3} with different values for parameter $r$}
\label{table6} \centering
\begin{tabular}{|lr|r|r|l|r|r|r|r|r|r}
 \hline
  Sample &  &Distribution&True &Size or &\multicolumn{5}{c|}{Probability of Rejecting $H_0$  for Tests}\\
     \cline{6-10}
  Size($n$)&$k_n$& under $H_0$&distribution &power& $\mathcal{R}$& $\mathcal{R}_0$& $\mathcal{R}_1$
  & $\mathcal{R}_3$& $\mathcal{R}_5$\\
    \hline
 100 & 100 & $r=$2 & $r=$2 &size& 0.0623 & 0.0607 & 0.0610 & 0.0547& 0.0509\\
     \cline{5-10}
 100 & 100 & 2 & 1 &power&0.0000 & 0.0416 & 0.7327 & 0.8545 & 0.8613 \\
 100 & 100 & 2 & 3 &power& 0.6877 & 0.2145 & 0.6008 & 0.7410 & 0.7559\\
 \cline{1-10}
 100 & 200 & $r=$2 & $r=$2 &size& 0.0609 & 0.0632 & 0.0597 & 0.0516 & 0.0520\\
    \cline{5-10}
 100 & 200 & 2 & 1 &power& 0.0000 & 0.0331 & 0.7878 & 0.8529 & 0.8688\\
100 & 200 & 2 & 3 &power& 0.7450 & 0.1899 & 0.6616 & 0.7526 & 0.7626\\
 \cline{1-10}
100 & 400 & 2 & 2 &size& 0.0584 & 0.0617 & 0.0517 & 0.0476 & 0.0468\\
100 & 400 & 2 & 1 &power& 0.0000 & 0.0246 & 0.8237 & 0.8714 & 0.8712\\
100 & 400 & 2 & 3 &power& 0.7837 & 0.1690 & 0.6973 & 0.7543 & 0.7611\\
\cline{1-10}
1000 & 300 & $r=$2 & $r=$2 &size&0.0598 & 0.0559 & 0.0601 & 0.0547 & 0.0518 \\
    \cline{5-10}
1000 & 300 & 2 & 1 &power& 0.0000 & 0.9779 & 1.0000 & 1.0000 & 1.0000\\
1000 & 300 & 2 & 3 &power& 0.9999 & 0.8164 & 0.9997 & 1.0000 & 1.0000\\
 \cline{1-10}
1000 & 1000 & $r=$2 & $r=$2 &size&0.0582 & 0.0602 & 0.0579 & 0.0540 & 0.0544 \\
    \cline{5-10}
1000 & 1000 & 2 & 1 &power& 0.0000 & 0.5604 & 1.0000 & 1.0000 & 1.0000\\
1000 & 1000 & 2 & 3 &power& 1.0000 & 0.5143 & 1.0000 & 1.0000 & 1.0000\\
 \cline{1-10}
1000 & 3000 & $r=$2 & $r=$2 &size& 0.0551 & 0.0597 & 0.0536 & 0.0515 & 0.0516\\
    \cline{5-10}
1000 & 3000 & 2 & 1 &power& 0.0000 & 0.1361 & 1.0000 & 1.0000 & 1.0000\\
1000 & 3000 & 2 & 3 &power& 1.0000 & 0.3219 & 1.0000 & 1.0000 & 1.0000\\
 \cline{1-10}
1000 & 10000 & $r=$2 & $r=$2 &size& 0.0511 & 0.0633 & 0.0520 & 0.0488 & 0.0487\\
    \cline{5-10}
1000 & 10000 & 2 & 1 &power& 0.0000 & 0.0384 & 1.0000 & 1.0000 & 1.0000\\
1000 & 10000 & 2 &3 &power& 1.0000 & 0.2211 & 1.0000 & 1.0000 & 1.0000\\
   \hline
\end{tabular}}
\end{table}

\vspace{20pt}

\section{A real data application}\label{app}

As an application, we study the winning numbers from the Minnesota
Lottery Game {\it Daily 3}. The game has been played for many years,
and a winning number consisting of three digits is drawn daily. The
three digits are drawn from digits $0, 1, \cdots, 9$. Minnesota
Lottery does not reveal how the three digits are selected in its
official website.  According to the State Lottery Report Card
(available at the site
\url{https://www.lotterypost.com/lottery-report-card.asp}), the
winning numbers for Minnesota {\it Daily 3} are drawn by computer
programs.

In Game {\it Daily 3},  there are $1000$ possible drawing outcomes.
We are interested in whether the drawing mechanism for {\it Daily 3}
is random, this is, whether all $1000$ possible outcomes are equally
likely. Recent winning numbers for this game can be found at the
Minnesota Lottery web site
\url{https://www.mnlottery.com/games/lotto_games/daily_3/winning_s/}.
As an application, we examine some early data from Game
\textit{Daily 3}. We have collected a total of $2919$ data points
from August 14, 1990 to August 13, 1998.  The winning numbers were
drawn every day except Christmas days in years 1990, 1991 and 1992.
This old dataset was obtained from the official Minnesota lottery
website but now it is no longer available.  One may find it from
\url{https://www.lotterypost.com/results}.

Since our null hypothesis is that $p_1=\cdots=p_{1000}=1/1000$, all
three test statistics, $\mathcal{X}_n^2$, $S_{n1}$, and
$S_{n1}+c|S_{n2}|$, are the same. We first apply all $2919$ data for
the test. The observed $\mathcal{X}^2_n$ is $978.5677$ with mean
$999$ and standard deviation $44.69$, and the standardized statistic
is $(978.6777-999)/44.69=-0.4572$, which has a p-value $0.6962$.  To
apply the new test $\overline{\mathcal{R}}_1$, we assign a value
$1.25$ to these $800$ weights $c_i$ to the cells associated to the
drawing numbers whose first digits are smaller than $8$.  Then we
have an observed value $-0.1774$ for
$\big(S_{n1}+|\overline{S}_{n2}|-(k_n-1)\big)/\sigma_n$, and the
corresponding p-value is $0.6927$. Therefore, at the $5\%$
significance level, we couldn't reject the null hypothesis and
conclude that the $1000$ possible winning numbers may be drawn with
equal probability.


Next, we test the hypothesis that $p_1=\cdots=p_{1000}=1/1000$ based
on each of eight periods. A period starts from August 14 in one year
and ends next August 13. Namely,  Period 1 is from August 14, 1990
to August 13, 1991, and Period 2 is from August 14, 1991 to August
13, 1993, so on.   For each of the first seven periods, both tests
result in large p-values.  For Period 8, that is, during Aug 14,
1997 to Aug 13, 1998,  the p-value for test $\mathcal{R}$ is
$0.00085$, and the p-value from test $\overline{\mathcal{R}}_1$ is
$0.002867$.
 We identify that the data during August 14,
1997 and August 13, 1998 seem highly abnormal. We find out that both
winning numbers $(3,7,5)$ and  $(4,4,8)$ appeared $4$ times, and
each of the other $11$ numbers appeared $3$ times.

Our second test procedure consists of $8$ individual tests. In order
to control the overall type I error at the $5\%$ level, we use the
Bonferroni inequality for multiple comparisons, that is, we reject
the null hypothesis that $p_1=\cdots=p_{1000}=1/1000$ at level
$0.05$ if any of the $8$ individuals tests is rejected at level
$0.05/8=0.00625$. Since both tests $\mathcal{R}$ and
$\overline{\mathcal{R}}_1$ have a p-value smaller than $0.00625$ for
Period 8, we conclude that the hypothesis of equiprobability can be
rejected at level $0.05$.

\section{Concluding remarks}

In this paper, we investigate the performance of Pearson's
goodness-of-fit test when both the number of cells and the sample
size go to infinity. Under quite general conditions, we  have
obtained the limiting distributions for Pearson's goodness-of-fit
test statistic and Zelterman's $D^2$ test statistic. By decomposing
Pearson's test statistic, we propose new test statistics so as to
overcome the bias problem of Pearson's chi-squared test. Our
simulation study indicates that test $\mathcal{R}_1$ is superior to
Pearson's goodness-of-fit test $\mathcal{R}$ in general.
$\mathcal{R}_1$  gains a much larger power than $\mathcal{R}$ in
most cases and is no longer biased. Our new test
$\overline{\mathcal{R}}_1$ is also much more powerful than
$\mathcal{R}$ for testing the equiprobable cells. For small and
moderate sample sizes, as pointed out by a referee, one can apply
permutation procedures (Drikvandi et al.~\cite{DKV2012}) or
bootstrap procedures (Nordhausen et al.~\cite{NOTV2017}) to conduct
the tests.

\section*{Supplementary material}

Proofs of the main results in the paper are given in the Supplement.

\section*{Acknowledgements}

The authors would like to thank the associate editor and three
referees for their constructive suggestions that have led to
improvement in the paper. Chang's research was supported in part by
the Major Research Plan of the National Natural Science Foundation
of China (91430108), the National Basic Research Program
(2012CB955804), the National Natural Science Foundation of China
(11771322), and the Major Program of Tianjin University of Finance
and Economics (ZD1302). The research of Deli Li was partially
supported by a grant from the Natural Sciences and Engineering
Research Council of Canada [grant no. RGPIN-2019-06065]. The
research of Yongcheng Qi was supported in part by NSF Grant
DMS-1916014.

\section*{Disclosure statement}

No potential conflict of interest was reported by the authors.

 \renewcommand{\thesection}{A}
 \numberwithin{equation}{section}

\newpage

\setcounter{page}{1}


\noindent{\color{blue} Supplement}

\vspace{30pt}

\noindent{\bf \Large Pearson's goodness-of-fit tests for sparse
distributions}

\vspace{20pt} \noindent{\bf Shuhua Chang$^{a,b}$, Deli Li$^c$,
Yongcheng Qi$^d$}

\vspace{10pt}

{\small \noindent $^a$Coordinated Innovation Center for Computable
Modeling in Management Science, Yango University, Fujian 350015,
China

\noindent $^b$Coordinated Innovation Center for Computable Modeling
in Management Science, Tianjin University of Finance and Economics,
Tianjin 300222, China
\\ Email: szhang@tjufe.edu.cn

\vspace{10pt}

\noindent $^c$Department of Mathematical Sciences, Lakehead
University Thunder Bay,  Ontario, Canada P7B 5E1.
\\Email: dli@lakeheadu.ca

\vspace{10pt}

\noindent $^d$Department of Mathematics and Statistics, University
of Minnesota Duluth, 1117 University Drive, Duluth, MN 55812, USA.
\\Email: yqi@d.umn.edu
}

\section*{Appendix: Proofs of the Main Results}

For each  $j\in \{1,\cdots, n\}$, define random variable $X_j=i$ if
$E_i$ occurs in the $j$-th trial of the experiment.  Then $X_j,
~1\le j\le n$, are independent and identically distributed random
variables with $P(X_j=i)=p_i$ for $1\le i\le k_n$, $1\le j\le n$.
Define
\[
\delta_{i,j}=I(X_j=i)-p_i ~~\mbox{ for } 1\le i\le k_n,~ 1\le j\le
n\] and set
\[
\Delta_{i, \ell}=\sum^{\ell}_{j=1}\delta_{i,j}~~\mbox{ for } 1\le
i\le k_n, ~~1\le \ell \le n.
\]
For convenience, set $\Delta_{i,0}=0$ for any $1\le i\le k_n$.

From now on, we use $\E(\cdot)$ to denote the expectation under the
null hypothesis that $P(E_i)=p_i$ for $1\le i\le k_n$. When an
alternative is specified as $H_1$:  $P(E_i)=p_i'$ for $1\le i\le
k_n$, where $(p_1', \cdots, p_{k_n}')\ne (p_1, \cdots, p_{k_n})$,
$\E(\cdot |H_1)$ denotes the conditional expectation under $H_1$.

We can easily verify the following equations:
\begin{equation}\label{fact1}
\E(\delta_{i,j})=0,~~~\E(\delta_{i,j}^2)=p_i(1-p_i);
\end{equation}
\begin{equation}\label{fact2}
\E(\delta_{i,j}\delta_{i',j'})=0 ~~\mbox{ if } j\ne j';
\end{equation}
\begin{equation}\label{fact3}
\E(\delta_{i,j}\delta_{i',j})= \left\{
  \begin{array}{ll}
    p_i(1-p_i), & \hbox{ if } i=i'; \\
    -p_ip_{i'}, & \hbox{ if } i\ne i',
  \end{array}
\right.
\end{equation}
\begin{equation}\label{fact4}
\sum_{i=1}^{k_n}\delta_{i,j}=0~~\mbox{ and
}~\sum_{i=1}^{k_n}\Delta_{i,j}=0,
\end{equation}
where $1\le i, i'\le k_n$, $1\le j, j'\le n$ in the above equations.

Since $\sum^\ell_{j=1}I(X_j=i)$ is the sum of $\ell$ independent
Bernoulli random variables, its distribution is binomial. We have
\begin{equation}\label{fact5}
\E(\Delta_{i,\ell}^2)=\ell p_i(1-p_i),~~~1\le \ell\le n.
\end{equation}
We can also verify that
\begin{equation}\label{fact6}
\E(\Delta_{i_1,\ell}\Delta_{i_2,\ell})=-\ell p_{i_1}p_{i_2},~~~ 1\le
i_1\ne i_2\le k_n,~~ 1\le \ell\le n.
\end{equation}
For each $i\in \{1,\cdots, k_n\}$,  we have
\begin{equation}\label{o-e}
o_i=\sum^n_{j=1}I(X_j=i)~~~\mbox{ and }
o_i-e_i=\sum^n_{j=1}\delta_{i,j}=\Delta_{i,n}.
\end{equation}
We also need the following expectations under the alternative $H_1$
\begin{equation}\label{fact8}
\E(\delta_{i,j}|H_1)=p_i'-p_i~~\mbox{ and
}~~\E(\delta_{i,j}\delta_{i,j'}|H_1)=(p_i'-p_i)^2
\end{equation}
for  $1\le i\le k_n$, $1\le j\ne j'\le n$.

\begin{lemma}\label{lem1} Let $S_{n1}$ and $S_{n2}$ be defined in \eqref{SS}. Then
\begin{equation}\label{S1}
S_{n1}=\frac1n\sum^{k_n}_{i=1}\frac{1}{p_i}\sum_{1\le j_1\ne j_2\le
n}\delta_{i,j_1}\delta_{i,j_2}+k_n-1
\end{equation}
and
\begin{equation}\label{S2}
S_{n2}=\frac1n\sum^n_{j=1}\sum^{k_n}_{i=1}\frac{\delta_{i,j}}{p_i}.
\end{equation}
Under the null hypothesis that $P(X_1=i)=p_i$ for $1\le i\le k_n$,
we have
\begin{equation}\label{H0mean}
\E (S_{n1}-(k_n-1))=0, ~~\E(S_{n2})=0;
\end{equation}
Under the alternative $H_1$:  $P(X_1=i)=p_i'$ for $1\le i\le k_n$,
we have
\begin{equation}\label{H1mean}
\E\big(S_{n1}-(k_n-1)|H_1\big)=(n-1)\sum^{k_n}_{i=1}\frac{(p_i'-p_i)^2}{p_i},~~\E(S_{n2}|H_1)=\sum^{k_n}_{i=1}\frac{p_i'-p_i}{p_i}.
\end{equation}
\end{lemma}

\noindent{\it Proof.} With the notations in the beginning of the
section, \eqref{S2} can be verified easily by using \eqref{o-e}. To
show \eqref{S1}, notice that
\begin{eqnarray}\label{decompostion}
\mathcal{X}^2_n&=&\sum^{k_n}_{i=1}\frac{(\sum^n_{j=1}\delta_{i,j})^2}{np_i}\nonumber\\
&=&\frac1n\sum^{k_n}_{i=1}\frac{1}{p_i}\sum^n_{j_1=1}\sum^n_{j_2=1}\delta_{i,j_1}\delta_{i,j_2}\nonumber\\
&=&\frac1n\sum^{k_n}_{i=1}\frac{1}{p_i}\sum_{1\le j_1\ne j_2\le
n}\delta_{i,j_1}\delta_{i,j_2}+\frac1n\sum^{k_n}_{i=1}\frac{1}{p_i}\sum^n_{j=1}\delta_{i,j}^2.
\end{eqnarray}
Since $\delta_{i,j}^2=I(X_j=i)-2I(X_j=i)p_i+p_i^2$,
$\sum^{k_n}_{i=1}I(X_j=i)=1$, and $\sum^{k_n}_{i=1}p_i=1$, we have
\begin{eqnarray*}
\frac1n\sum^{k_n}_{i=1}\frac{1}{p_i}\sum^n_{j=1}\delta_{i,j}^2
&=&\frac1n\sum^{k_n}_{i=1}\sum^n_{j=1}\Big(\frac{I(X_j=i)}{p_i}-2I(X_j=i)+p_i\Big)\\
&=&\frac1n\sum^n_{j=1}\sum^{k_n}_{i=1}\Big(\frac{I(X_j=i)}{p_i}-2I(X_j=i)+p_i\Big)\\
&=&\frac1n\sum^n_{j=1}\Big(\sum^{k_n}_{i=1}\frac{I(X_j=i)}{p_i}-1\Big)\\
&=&\frac1n\sum^n_{j=1}\sum^{k_n}_{i=1}\frac{I(X_j=i)-p_i}{p_i}+(k_n-1)\\
&=&\frac1n\sum^n_{j=1}\sum^{k_n}_{i=1}\frac{\delta_{i,j}}{p_i}+(k_n-1)\\
&=&S_{n2}+k_n-1,
\end{eqnarray*}
which, together with \eqref{decompostion}, yields \eqref{S1}.

Equations \eqref{H0mean} and \eqref{H1mean} follow from \eqref{S1},
\eqref{S2}, \eqref{fact2} and \eqref{fact8}. This completes the
proof. \qed

\begin{lemma}\label{lem2} Let $c_i$, $1\le i\le k_n$, be non-negative numbers with $\sum^{k_n}_{i=1}c_i=k_n$.  Then for any $j\ge  1$
\begin{equation}\label{jj}
\beta_{nj} :=\sum^{k_n}_{i=1}\frac{c_i^{j+1}}{p_i^j}-k_n^{j+1}\ge 0,
\end{equation}
and the equality holds only if for some $c>0$, $c_i=cp_i$ for $1\le
i\le k_n$.
\end{lemma}

\noindent{\it Proof.}  We will prove \eqref{jj} by induction.  By
using the Cauchy-Schwarz inequality we get
\[
\sum^{k_n}_{i=1}\frac{c_i^2}{p_i}=\sum^{k_n}_{i=1}\big(\frac{c_i}{p_i^{1/2}}\big)^2\sum^{k_n}_{i=1}(p_i^{1/2})^2\ge
\Big(\sum^{k_n}_{i=1}\frac{c_i}{p_i^{1/2}}p_i^{1/2}\Big)^2=k_n^2
\]
and the equality holds only if $(\frac{c_1}{\sqrt{p_1}},\cdots,
\frac{c_{k_n}}{\sqrt{p_{k_n}}})=c(\sqrt{p_1}, \cdots,
\sqrt{p_{k_n}})$ for some $c>0$, and the latter is  equivalent to
$c_i=cp_i$ for $1\le i\le k_n$. This implies \eqref{jj} holds with
$j=1$. Now assume \eqref{jj} holds for all $j\le j_0$ for some
$j_0\ge 1$. We need to show \eqref{jj} holds with $j=j_0+1$. If
$j_0+1=2k$ is an even number where $k\ge 1$, then it follows from
the Cauchy-Schwarz inequality that
\[
\sum^{k_n}_{i=1}\frac{c_i^{j_0+2}}{p_i^{j_0+1}}=
\frac{1}{k_n}\sum^{k_n}_{i=1}\big(\frac{c_i^{k+1/2}}{p_i^k}\big)^2\sum^{k_n}_{i=1}
(c_i^{1/2})^2\ge
\frac{1}{k_n}\Big(\sum^{k_n}_{i=1}\frac{c_i^{k+1}}{p_i^k}\Big)^2\ge
\frac{1}{k_n}(k_n^{k+1})^2=k_n^{j_0+2},
\]
 and the equality holds only if $c_i=cp_i$ for $1\le
i\le k_n$ for some $c>0$, proving \eqref{jj} with $j=j_0+1$. If
$j_0+1=2k+1$ is an odd number with $k\ge 1$,
 then again from the Cauchy-Schwarz inequality
\begin{eqnarray*}
\sum^{k_n}_{i=1}\frac{c_i^{j_0+2}}{p_i^{j_0+1}}&=&\sum^{k_n}_{i=1}\big(\frac{c^{k+1}}{p_i^{(2k+1)/2}}\big)^2\sum^{k_n}_{i=1}(p_i^{1/2})^2\ge
\Big(\sum^{k_n}_{i=1}\frac{c_i^{k+1}p_i^{1/2}}{p_i^{(2k+1)/2}}\Big)^2\\
&\ge& \Big(\sum^{k_n}_{i=1}\frac{c_i^{k+1}}{p_i^{k}}\Big)^2\ge
\big(k_n^{k+1}\big)^2 = k_n^{j_0+2},
\end{eqnarray*}
i.e.  \eqref{jj} holds with $j=j_0+1$. Similarly, we have the
equality only if $c_i=cp_i$ for $1\le i\le k_n$ for some $c>0$. This
completes the proof. \qed

\begin{lemma}\label{lem3} Assume $\{n_r,~ r\ge 1\}$ is an increasing sequence of positive integers.  If \eqref{C3}
holds with $n=n_r$ as $r\to\infty$, then \eqref{CLT2} holds with
$n=n_r$ as $r\to\infty$.
\end{lemma}

\noindent{\it Proof.} For brevity, we will drop the subscript $r$
and write $n_r$ as $n$ in the proof.

To prove \eqref{CLT2}, we will employ a martingale technique.  To
this end,  we first rewrite $S_{n1}$ as
\begin{eqnarray*}
S_{n1}-(k_n-1)&=&\frac1n\sum^{k_n}_{i=1}\frac{1}{p_i}\sum_{1\le j_1\ne j_2\le n}\delta_{i,j_1}\delta_{i,j_2}\\
&=&\frac2n\sum^{k_n}_{i=1}\frac{1}{p_i}\sum_{1\le j_1< j_2\le n}\delta_{i,j_1}\delta_{i,j_2}\\
&=&\frac2n\sum^{k_n}_{i=1}\frac{1}{p_i}\sum^n_{\ell=2}\sum_{j=1}^{\ell-1}\delta_{i,j}\delta_{i,\ell}\\
&=&\sum^n_{\ell=2}\Big(\frac2n\sum^{k_n}_{i=1}\frac{1}{p_i}\sum_{j=1}^{\ell-1}\delta_{i,j}\delta_{i,\ell}\Big)\\
&=&\sum^n_{\ell=2}\Big(\frac2n\sum^{k_n}_{i=1}\frac{1}{p_i}\Delta_{i,\ell-1}\delta_{i,\ell}\Big).
\end{eqnarray*}

Let $\mathcal{F}_{n\ell}=\sigma(X_1, X_2, \cdots, X_{\ell})$ denote
the $\sigma$-algebra generated by $\{X_1, X_2, \cdots, X_{\ell}\}$
for $1\le\ell\le n$, and $\mathcal{F}_{n0}=\{\phi, \Omega\}$ is the
trivial $\sigma$-algebra. Now set
\begin{equation}\label{znl}
z_{n\ell}=\frac2n\sum^{k_n}_{i=1}\frac{1}{p_i}\Delta_{i,\ell-1}\delta_{i,\ell},~~1\le
\ell\le n.
\end{equation}
Note that $z_{n1}=0$.  By the independence of $\delta_{i,\ell}$ and
$\mathcal{F}_{n(\ell-1)}$, we have
\[
\E(z_{n\ell}|\mathcal{F}_{n(\ell-1)})
=\frac2n\sum^{k_n}_{i=1}\frac{1}{p_i}\Delta_{i,\ell-1}\E(\delta_{i,\ell}|\mathcal{F}_{n(\ell-1)})=0
~~~\mbox{ for } 1\le \ell\le n.
\]
Therefore, $\{z_{n\ell}, ~\mathcal{F}_{n\ell}, ~ 1\le \ell\le
n,~n\ge 2\}$ form an array of martingale differences.  Since
$S_{n1}=\sum^n_{\ell=1}z_{n\ell}$, it is sufficient to show that
\begin{equation}\label{mCLT}
\frac{\sum^{p_n}_{\ell=1}z_{n\ell}}{\sigma_{n1}}\td N(0,1).
\end{equation}
In view of Corollary 3.1 in Hall and Heyde~\cite{Hall}, the
martingale central limit theorem \eqref{mCLT} holds if the following
two conditions hold:
\begin{equation}\label{martingale1}
\frac{1}{\sigma_{n1}^2}\sum^n_{\ell=1}\E(z_{n\ell}^2I(|z_{n\ell}|\ge
\varepsilon \sigma_{n1})|\mathcal{F}_{n(\ell-1)})\to 0~~~\mbox{ in
probability}
\end{equation}
for every $\varepsilon>0$, and
\begin{equation}\label{martingale2}
\frac{1}{\sigma_{n1}^2}\sum^{n}_{\ell=1}\E(z_{n\ell}^2|\mathcal{F}_{n(\ell-1)})\to
1~~~\mbox{ in probability}.
\end{equation}

Recall $z_{n\ell}$ is defined in \eqref{znl}.  We have
\begin{equation}\label{znlsquare}
z_{n\ell}^2=\frac4{n^2}\sum_{1\le i_1, i_2\le
k_n}\frac{\Delta_{i_1,\ell-1}\Delta_{i_2,\ell-1}}{p_{i_1}p_{i_2}}\delta_{i_1,\ell}\delta_{i_2,\ell}.
\end{equation}
By taking conditional expectations on $\mathcal{F}_{n(\ell-1)}$,
using the independence of $\delta_{i_1,\ell}\delta_{i_2,\ell}$ and
$\mathcal{F}_{n(\ell-1)}$ we get
\begin{eqnarray*}
&&\E(z_{n\ell}^2|\mathcal{F}_{n(\ell-1)})\\
&=&\frac4{n^2}\sum_{1\le i_1, i_2\le k_n}\frac{\Delta_{i_1,\ell-1}\Delta_{i_2,\ell-1}}{p_{i_1}p_{i_2}}\E\big(\delta_{i_1,\ell}\delta_{i_2,\ell}|\mathcal{F}_{n(\ell-1)}\big)\\
&=&\frac4{n^2}\sum_{1\le i_1, i_2\le k_n}\frac{\Delta_{i_1,\ell-1}\Delta_{i_2,\ell-1}}{p_{i_1}p_{i_2}}\E\big(\delta_{i_1,\ell}\delta_{i_2,\ell}\big)\\
&=&\frac4{n^2}\Big(\sum_{1\le i_1=i_2\le
k_n}\frac{\Delta_{i_1,\ell-1}\Delta_{i_2,\ell-1}}{p_{i_1}p_{i_2}}\E\big(\delta_{i_1,\ell}\delta_{i_2,\ell}\big)
+\sum_{1\le i_1\ne i_2\le
k_n}\frac{\Delta_{i_1,\ell-1}\Delta_{i_2,\ell-1}}{p_{i_1}p_{i_2}}\E\big(\delta_{i_1,\ell}\delta_{i_2,\ell}\big)
\Big).
\end{eqnarray*}
In view of \eqref{fact3} and \eqref{fact4}, we get for $2\le \ell\le
n$
\begin{eqnarray*}
&&\E(z_{n\ell}^2|\mathcal{F}_{n(\ell-1)})\\
&=&\frac4{n^2}\Big(\sum_{1\le i\le
k_n}\frac{\Delta_{i,\ell-1}^2}{p_{i}}(1-p_i) -\sum_{1\le i_1\ne
i_2\le k_n}\Delta_{i_1,\ell-1}\Delta_{i_2,\ell-1}
\Big)\\
&=&\frac4{n^2}\Big(\sum_{1\le i\le
k_n}\frac{\Delta_{i,\ell-1}^2}{p_{i}}(1-p_i)+\sum_{1\le i_1=i_2\le
k_n}\Delta_{i_1,\ell-1}\Delta_{i_2,\ell-1} -\sum_{1\le i_1,i_2\le
k_n}\Delta_{i_1,\ell-1}\Delta_{i_2,\ell-1}
\Big)\\
&=&\frac4{n^2}\Big(\sum_{1\le i\le
k_n}\frac{\Delta_{i,\ell-1}^2}{p_{i}}(1-p_i)+\sum_{1\le i\le
k_n}\Delta_{i,\ell-1}^2
-(\sum_{1\le i\le n}\Delta_{i,\ell-1}\big)^2 \Big)\\
&=&\frac4{n^2}\sum_{1\le i\le k_n}\frac{\Delta_{i,\ell-1}^2}{p_{i}}.
\end{eqnarray*}
Therefore, we get the conditional variance for the martingale
differences $\{z_{n\ell}, ~\mathcal{F}_{n\ell}, ~ 1\le \ell\le
n,~n\ge 2\}$
\begin{equation}\label{sigman|c}
\sigma^2_{n|c}:=\sum^n_{\ell=2}\E(z_{n\ell}^2|\mathcal{F}_{n(\ell-1)})=\frac4{n^2}\sum^n_{\ell=2}
\sum_{1\le i\le k_n}\frac{\Delta_{i,\ell-1}^2}{p_{i}},
\end{equation}
and from \eqref{fact5}
\begin{eqnarray}
\E(\sigma^2_{n|c})
&=&\frac4{n^2}\sum^n_{\ell=2}\sum_{1\le i\le k_n}\frac{(\ell-1)p_i(1-p_i)}{p_i}\nonumber\\
&=&\frac4{n^2}\sum^n_{\ell=2}(\ell-1)\sum_{1\le i\le k_n}(1-p_i)\nonumber\\
&=&\frac4{n^2}\frac{n(n-1)}{2}(k_n-1)\nonumber\\
&=&\frac{2(n-1)(k_n-1)}{n}\nonumber\\
&=&\sigma_{n1}^2, \label{Esigman|c}
\end{eqnarray}
where $\sigma_{n1}^2$ is the variance defined in Theorem~\ref{thm1}.

Taking into account the above computation, \eqref{martingale1} and
\eqref{martingale2} follow if we can verify the following equations
\begin{equation}\label{4thmoment}
\sum^n_{\ell=1}\E(z^4_{n\ell})=o(\sigma_{n1}^4)~~\mbox{ as
}n\to\infty
\end{equation}
and
\begin{equation}\label{2ndmoment}
\E(\sigma^2_{n|c}-\sigma_{n1}^2)^2=o(\sigma_{n1}^4)~~\mbox{ as
}n\to\infty.
\end{equation}

We will prove \eqref{2ndmoment} first.  Rewrite
\begin{eqnarray*}
\sigma^2_{n|c}
&=&\frac4{n^2}\sum^n_{\ell=2}\sum_{1\le i\le k_n}\frac{\Delta_{i,\ell-1}^2}{p_{i}}\\
&=&\frac4{n^2}\sum^n_{\ell=2}\sum_{i=1}^{k_n}\frac{1}{p_i}\sum_{1\le j_1,j_2\le \ell-1}\delta_{i,j_1}\delta_{i,j_2}\\
&=&\frac4{n^2}\sum^n_{\ell=2}\sum_{i=1}^{k_n}\frac{1}{p_i}
\Big(\sum_{j=1}^{k_n}\delta_{i,j}^2+2\sum_{1\le j_1<j_2\le \ell-1}\delta_{i,j_1}\delta_{i,j_2}\Big)\\
&=&\frac4{n^2}\sum^n_{\ell=2}\sum_{i=1}^{k_n}\frac{1}{p_i}\sum_{j=1}^{\ell-1}\delta_{i,j}^2+
+\frac8{n^2}\sum^n_{\ell=3}\sum_{i=1}^{k_n}\frac{1}{p_i}\sum_{1\le j_1<j_2\le \ell-1}\delta_{i,j_1}\delta_{i,j_2}\\
&=&:I_{n1}+I_{n2}.
\end{eqnarray*}
Then \eqref{2ndmoment} follows if
\begin{equation}\label{I1&I2}
\E\big(I_{n1}-\sigma_{n1}^2\big)^2=o(\sigma_{n1}^4)~~\mbox { and
}~~\E\big(I_{n2}^2\big)=o(\sigma_{n1}^4).
\end{equation}

Note that
\begin{eqnarray*}
I_{n1}
&=&\frac4{n^2}\sum^n_{\ell=2}\sum_{j=1}^{\ell-1}\sum_{i=1}^{k_n}\frac{1}{p_i}\delta_{i,j}^2\\
&=&\frac4{n^2}\sum^n_{\ell=2}\sum_{j=1}^{\ell-1}\sum_{i=1}^{k_n}\big(\frac{I(X_j=i)}{p_i}+p_i-2I(X_j=i)\big)\\
&=&\frac4{n^2}\sum^n_{\ell=2}\sum_{j=1}^{\ell-1}\Big(\sum_{i=1}^{k_n}\frac{I(X_j=i)}{p_i}-1\Big)\\
&=&\frac4{n^2}\sum^n_{\ell=2}\sum_{j=1}^{\ell-1}\Big(\sum_{i=1}^{k_n}\frac{I(X_j=i)-p_i}{p_i}+k_n-1\Big)\\
&=&\frac4{n^2}\sum^n_{\ell=2}\sum_{j=1}^{\ell-1}\Big(\sum_{i=1}^{k_n}\frac{\delta_{i,j}}{p_i}+k_n-1\Big)\\
&=&\frac4{n^2}\sum^n_{\ell=2}\sum_{j=1}^{\ell-1}\sum_{i=1}^{k_n}\frac{\delta_{i,j}}{p_i}+\frac4{n^2}\sum^n_{\ell=2}\sum_{j=1}^{\ell-1}(k_n-1)\\
&=&\frac4{n^2}\sum^n_{\ell=2}\sum_{j=1}^{\ell-1}\sum_{i=1}^{k_n}\frac{\delta_{i,j}}{p_i}+\sigma_{n1}^2\\
&=&\frac4{n^2}\sum^{n-1}_{j=1}\sum_{i=1}^{k_n}\frac{(n-j)\delta_{i,j}}{p_i}+\sigma_{n1}^2.
\end{eqnarray*}
The last step is obtained from the previous one by taking summation
over $\ell$ first.   In view of \eqref{fact2} and \eqref{fact3} we
get
\begin{eqnarray*}
\E\big(I_{n1}-\sigma_{n1}^2\big)^2
&=&\frac{16}{n^4}\E\big( \sum^{n-1}_{j=1}\sum_{i=1}^{k_n}\frac{(n-j)\delta_{i,j}}{p_i}\big)^2\\
&=&\frac{16}{n^4}\sum_{1\le j_1,j_2\le n-1}\sum_{1\le i_1, i_2\le k_n}\frac{(n-j_1)(n-j_2)\E\big(\delta_{i_1,j_1}\delta_{i_2,j_2}\big)}{p_{i_1}p_{i_2}}\\
&=&\frac{16}{n^4}\sum_{j=1}^{n-1}\sum_{1\le i_1, i_2\le k_n}\frac{(n-j)^2\E\big(\delta_{i_1,j}\delta_{i_2,j}\big)}{p_{i_1}p_{i_2}}\\
&=&\frac{16}{n^4}\sum_{j=1}^{n-1}(n-j)^2\Big(\sum_{1\le i \le
k_n}\frac{p_i(1-p_i)}{p_{i}^2} +
\sum_{1\le i_1\ne i_2\le k_n}\frac{-p_{i_1}p_{i_2}}{p_{i_1}p_{i_2}}\Big)\\
&=&\frac{16}{n^4}\sum_{j=1}^{n-1}(n-j)^2\Big(\sum_{i=1}^{k_n}\frac{1}{p_i}-k_n^2\Big)\\
&=&\frac{O(k_n^2)}{nk_n^2}\Big(\sum_{i=1}^{k_n}\frac{1}{p_i}-k_n^2\Big)\\
&=&o(\sigma_{n1}^4).
\end{eqnarray*}
We have used the fact that
$\lim_{n\to\infty}\frac{1}{nk_n^2}(\sum_{i=1}^{k_n}\frac{1}{p_i}-k_n^2)=0$,
which follows from condition \eqref{C3} since
\begin{equation}\label{sonofC3}
\sum_{i=1}^{k_n}\frac{1}{p_i}\le
\sqrt{\sum^{k_n}_{i=1}\frac{1}{p_i^2}\sum^{k_n}_{i=1}1}=o(nk_n)k_n^{1/2}=o(nk_n^2)
\end{equation}
from the Cauchy-Schwarz inequality. The first part of \eqref{I1&I2}
is obtained.

To prove the second part of \eqref{I1&I2}, we can take summation
over $\ell$ first. Then we have
\begin{eqnarray*}
I_{n2}
&=&\frac8{n^2}\sum^n_{\ell=3}\sum_{i=1}^{k_n}\frac{1}{p_i}\sum_{1\le j_1<j_2\le \ell-1}\delta_{i,j_1}\delta_{i,j_2}\\
&=&\frac8{n^2}\sum^{k_n}_{j=2}\sum_{i=1}^{k_n}\frac{n-j}{p_i}\Delta_{i,
j-1}\delta_{i,j}.
\end{eqnarray*}
We note that $\E\big(\Delta_{i_1, j_1-1}\delta_{i_1,j_1}\Delta_{i_2,
j_2-1}\delta_{i_2,j_2} \big)=0$ if $j_1\ne j_2$ for any $1\le i_1,
i_2\le k_n$. We thus have from equations \eqref{fact1},
\eqref{fact3}, \eqref{fact5} and \eqref{fact6} that
 \begin{eqnarray*}
\E(I_{n2}^2) &=&\frac{64}{n^4}\sum_{2\le j_1, j_2\le n}\sum_{1\le
i_1, i_2\le k_n}\frac{(n-j_1)(n-j_2)}{p_{i_1}p_{i_2}}
\E\big(\Delta_{i_1, j_1-1}\delta_{i_1,j_1}\Delta_{i_2, j_2-1}\delta_{i_2,j_2}\big)\\
&=&\frac{64}{n^4}\sum_{2\le j\le n}\sum_{1\le i_1, i_2\le
k_n}\frac{(n-j)^2}{p_{i_1}p_{i_2}}
\E\big(\Delta_{i_1, j-1}\delta_{i_1,j}\Delta_{i_2, j-1}\delta_{i_2,j}\big)\\
&=&\frac{64}{n^4}\sum_{2\le j\le n}\sum_{1\le i_1, i_2\le
k_n}\frac{(n-j)^2}{p_{i_1}p_{i_2}}
\E\big(\Delta_{i_1, j-1}\Delta_{i_2, j-1}\big)\E\big(\delta_{i_1,j}\delta_{i_2,j}\big)\\
&=&\frac{64}{n^4}\sum_{2\le j\le n}(n-j)^2\Big(\sum_{1\le i\le
k_n}\frac{(j-1)p_i^2(1-p_i)^2}{p_i^2} +\sum_{1\le i_1\ne i_2\le
k_n}\frac{(j-1)p_{i_1}^2p_{i_2}^2}{p_{i_1}p_{i_2}}
\Big)\\
&=&\frac{64}{n^4}\sum_{2\le j\le n}(n-j)^2(j-1)\Big(\sum_{1\le i\le
k_n}(1-p_i)^2+\sum_{1\le i_1\ne i_2\le k_n}p_{i_1}p_{i_2}
\Big)\\
&=&\frac{64(k_n-1)}{n^4}\sum_{2\le j\le n}(n-j)^2(j-1)\\
&=&O(k_n-1)\\
&=&o(\sigma_{n1}^4),
\end{eqnarray*}
proving the second part of \eqref{I1&I2}.

Finally, we show \eqref{4thmoment}. To estimate $\E(z_{n\ell}^4)$,
we need the following calculations which are straightforward:
\begin{eqnarray*}
d_4(i):&=&\E(\delta_{i,\ell}^4)=p_i(1-p_i)^4+p_i^4(1-p_i),\\
d_{3,1}(i,j):&=&\E(\delta_{i,\ell}^3\delta_{j,\ell})= p_ip_j(1-p_i-p_j)-(1-p_i)^3p_ip_j- p_i^3(1-p_j)p_j,\\
d_{2,2}(i,j):&=&\E(\delta_{i,\ell}^2\delta_{j,\ell}^2)=p_ip_j(1-p_i-p_j)+p_ip_j^2(1-p_i)^2+p_i^2p_j(1-p_j)^2,\\
d_{2,1,1}(i,j,m):&=&\E(\delta_{i,\ell}^2\delta_{j,\ell}\delta_{m,\ell})=p_i^2p_jp_m(1-p_i-p_j-p_m)+(1-p_i)^2p_ip_jp_m \\
&&~~-p_i^2p_j(1-p_j)p_m-p_i^2p_jp_m(1-p_m),\\
d_{1,1,1,1}(i,j,m,r):&=&\E(\delta_{i,\ell}\delta_{j,\ell}\delta_{m,\ell}\delta_{r,\ell})\\
                      &=&p_ip_jp_mp_r(1-p_i-p_j-p_m-p_r)-p_ip_jp_mp_r(1-p_i)\\
&&~~-p_ip_jp_mp_r(1-p_j)-p_ip_jp_mp_r(1-p_m)-p_ip_jp_mp_r(1-p_r)\\
&=&-3p_ip_jp_mp_r,
\end{eqnarray*}
where integers $i, j, m, r\in \{1, \cdots, k_n\}$ assume different
values if they appear in the same equations. Then it follows from
the above equations that
\[
d_4(i)\le 2p_i,~|d_{3,1}(i,j)|\le 3p_ip_j,~d_{2,2}(i,j)\le
3p_ip_j,~~|d_{2,1,1}(i,j,m)|\le 4p_ip_jp_m
\]
and
\[
d_{1,1,1,1}(i,j,m,r)=-3p_ip_jp_mp_r.
\]

Now we estimate
$\E\big(\Delta_{i_1,\ell}\Delta_{i_2,\ell}\Delta_{i_3,\ell}\Delta_{i_4,\ell}\big)$.
Note that
\begin{eqnarray*}
\E\big(\Delta_{i_1,\ell}\Delta_{i_2,\ell}\Delta_{i_3,\ell}\Delta_{i_4,\ell}\big)
&=&\E\big(\sum^\ell_{\ell_1=1}\delta_{i_1,\ell_1}\sum^\ell_{\ell_2=1}\delta_{i_2,\ell_2}\sum^\ell_{\ell_3=1}\delta_{i_3,\ell_3}
\sum^\ell_{\ell_4=1}\delta_{i_4,\ell_4}\big)\\
&=&\E\big(\sum_{1\le \ell_1, \ell_2, \ell_3,\ell_4\le \ell}\delta_{i_1,\ell_1}\delta_{i_2,\ell_2}\delta_{i_3,\ell_3}\delta_{i_4,\ell_4}\big)\\
&=&\sum_{1\le \ell_1, \ell_2, \ell_3,\ell_4\le
\ell}\E\big(\delta_{i_1,\ell_1}\delta_{i_2,\ell_2}\delta_{i_3,\ell_3}\delta_{i_4,\ell_4}\big).
\end{eqnarray*}
It is easy to see that
$\E\big(\delta_{i_1,\ell_1}\delta_{i_2,\ell_2}\delta_{i_3,\ell_3}\delta_{i_4,\ell_4}\big)\ne
0$ only if $\ell_1=\ell_2=\ell_3=\ell_4$, or $\ell_1, \ell_2,
\ell_3, \ell_4$ form two distinct matching pairs such as
$\ell_1=\ell_2\ne \ell_3=\ell_4$.  Therefore, we have
\begin{eqnarray*}
\E\big(\Delta_{i_1,\ell}\Delta_{i_2,\ell}\Delta_{i_3,\ell}\Delta_{i_4,\ell}\big)
&=&\sum_{1\le j \le \ell}\E\big(\delta_{i_1,j}\delta_{i_2,j}\delta_{i_3,j}\delta_{i_4,j}\big)+\sum_{1\le m\ne r \le \ell}\E\big(\delta_{i_1,m}\delta_{i_2,m}\big)\E\big(\delta_{i_3,r}\delta_{i_4,r}\big)\\
&&\hspace{-55pt}+\sum_{1\le m\ne r \le
j}\E\big(\delta_{i_1,m}\delta_{i_3,m}\big)\E\big(\delta_{i_2,r}\delta_{i_4,r}\big)+
\sum_{1\le m\ne r \le
j}\E\big(\delta_{i_1,m}\delta_{i_4,m}\big)\E\big(\delta_{i_2,r}\delta_{i_3,r}\big).
\end{eqnarray*}
This, together with \eqref{fact3}, yields
\begin{eqnarray*}
D_4^{(\ell)}(i):&=&\E(\Delta_{i,\ell}^4)=\ell d_4(i)+3\ell(\ell-1)p_i^2(1-p_i)^2,\\
D_{3,1}^{(\ell)}(i,j):&=&\E(\Delta_{i,\ell}^3\Delta_{j,\ell})=\ell d_{3,1}(i,j)-3\ell(\ell-1)p_i^2(1-p_i)p_j, \\
D_{2,2}^{(\ell)}(i,j):&=&\E(\Delta_{i,\ell}^2\Delta_{j,\ell}^2)= \ell d_{2,2}(i,j)+\ell(\ell-1)\big(p_ip_j(1-p_i)(1-p_j)+2p_i^2p_j^2\big),\\
D_{2,1,1}^{(\ell)}(i,j,m):&=&\E(\Delta_{i,\ell}^2\Delta_{j,\ell}\Delta_{m,\ell})=\ell d_{2,1,1}(i,j,m)+\ell(\ell-1)(  3p_i^2p_jp_m-p_ip_jp_m),\\
D_{1,1,1,1}^{(\ell)}(i,j,m,r):&=&\E(\Delta_{i,\ell}\Delta_{j,\ell}\Delta_{m,\ell}\Delta_{r,\ell})=\ell
d_{1,1,1,1}(i,j, m,r)+3\ell(\ell-1)p_ip_jp_mp_r,
\end{eqnarray*}
where $i, j, m, r$ are different integers if they appear in the same
equation.  Therefore,  we get $2\le \ell\le n$
\begin{eqnarray*}
D_4^{(\ell-1)}(i)d_4(i)&\le &4\ell p_i^2+6\ell^2p_i^3,\\
D_{3,1}^{(\ell-1)}(i,j)d_{3,1}(i,j)&\le &9\ell p_i^2p_j^2+9\ell^2p_i^3p_j^2, \\
D_{2,2}^{(\ell-1)}(i,j)d_{2,2}(i,j)&\le &9\ell^2 p_i^2p_j^2,\\
D_{2,1,1}^{(\ell-1)}(i,j,m)d_{2,1,1}(i,j,m)&\le &16\ell^2p_i^2p_j^2p_m^2,\\
D_{1,1,1,1}^{(\ell-1)}(i,j,m,r)d_{1,1,1,1}(i,j,m,r)&\le
&9\ell^2p_i^2p_j^2p_m^2p_r^2.
\end{eqnarray*}

It follows from \eqref{znlsquare} that for $2\le \ell\le n$
\[
z_{n\ell}^4=\frac4{n^4}\sum_{1\le i_1, i_2, i_3, i_4\le
k_n}\frac{\Delta_{i_1,\ell-1}\Delta_{i_2,\ell-1}\Delta_{i_3,\ell-1}\Delta_{i_4,\ell-1}}
{p_{i_1}p_{i_2}p_{i_3}p_{i_4}}\delta_{i_1,\ell}\delta_{i_2,\ell}\delta_{i_3,\ell}\delta_{i_4,\ell},
\]
and thus
\begin{equation}\label{EZ}
\E(z_{n\ell}^4)=\frac4{n^4}\sum_{1\le i_1, i_2, i_3, i_4\le
k_n}\frac{\E\big(\Delta_{i_1,\ell-1}\Delta_{i_2,\ell-1}\Delta_{i_3,\ell-1}\Delta_{i_4,\ell-1}\big)}
{p_{i_1}p_{i_2}p_{i_3}p_{i_4}}\E\big(\delta_{i_1,\ell}\delta_{i_2,\ell}\delta_{i_3,\ell}\delta_{i_4,\ell}\big).
\end{equation}
We will divide $\{(i_1, i_2, i_3, i_4): 1\le i_1, i_2, i_3, i_4\le
n\}$ into several subsets, and classify these subsets into groups.
The contributions to $\E(z_{n\ell}^4)$ from subsets within each
group are the same, and we will list only one representative subset
within each group. The above inequalities will be used in the
following estimations.
\begin{itemize}
  \item \noindent{Group 1}:  $i_1, i_2, i_3, i_4$ are the same, that is, $\{(i_1, i_2, i_3, i_4): 1\le i_1=i_2=i_3=i_4\le n \}=:G_1$.
The sum of the summands over $G_1$ on the right-hand side of
\eqref{EZ} is equal to
\[
\sigma_1^{(\ell)}:=\sum_{1\le i\le
n}\frac{1}{p_i^4}D_4^{(\ell-1)}(i)d_4(i)\le
4(\ell-1)\sum^{k_n}_{i=1}\frac{1}{p_i^2}+6(\ell-1)^2\sum^{k_n}_{i=1}\frac{1}{p_i}.
\]

  \item \noindent{Group 2}:  Exactly three of $i_1, i_2, i_3, i_4$ are the same. A representative is $\{(i_1, i_2, i_3, i_4): 1\le i_1=i_2=i_3\ne i_4\le n\}=:G_2$. There are 4 such subsets. The sum of the summands over $G_2$ on the right-hand side of \eqref{EZ} is equal to
\[
\sigma_2^{(\ell)}:=\sum_{1\le i, \ne j\le
n}\frac{1}{p_i^3p_j}D_{3,1}^{(\ell-1)}(i,j)d_{3,1}(i,j)\le
9(\ell-1)\sum^{k_n}_{i=1}\frac{1}{p_i}+9(\ell-1)^2k_n.
\]

  \item \noindent{Group 3}:  $i_1, i_2, i_3, i_4$ form two distinct matching pairs.
  A representative is $\{(i_1, i_2, i_3, i_4): 1\le i_1=i_2\ne i_3=i_4\le
  n\}=:G_3$.
  There are $3$ such subsets within this group. The sum of the summands over $G_3$ on the right-hand side of \eqref{EZ} is equal to
\[
\sigma_3^{(\ell)}:=\sum_{1\le i\ne j\le
n}\frac{1}{p_i^2p_j^2}D_{2,2}^{(\ell-1)}(i,j)d_{2,2}(i,j)\le
9(\ell-1)^2k_n^2.
\]

  \item \noindent{Group 4}: Exactly two of $i_1, i_2, i_3, i_4$ are the same and there is only one matching pair.  A representative is
$\{(i_1, i_2, i_3, i_4): 1\le i_1=i_2 \ne i_3 \ne i_4\le n\}=G_4$.
There are $6$ subsets within this group.  The sum of the summands
over $G_4$ on the right-hand side of \eqref{EZ} is equal to
\[
\sigma_4^{(\ell)}:=\sum_{1\le i\ne j\ne m\le
n}\frac{1}{p_i^2p_jp_m}D_{2,1,1}^{(\ell-1)}(i,j,m)d_{2,1,1}(i,j,m)\le
16(\ell-1)^2k_n.
\]

  \item \noindent{Group 5}: $i_1, i_2, i_3, i_4$ are distinct, that is, $\{(i_1, i_2, i_3, i_4): 1\le i_1\ne i_2\ne i_3\ne i_4\le n\}=:G_5$.
The sum of the summands over $G_5$ on the right-hand side of
\eqref{EZ} is equal to
\[
\sigma_5^{(\ell)}:=\sum_{1\le i\ne j\ne m\ne r\le
n}\frac{1}{p_ip_jp_mp_r}D_{1,1,1,1}^{(\ell-1)}(i,j,m,
r)d_{1,1,1,1}(i,j,m,r)\le 9(\ell-1)^2.
\]
\end{itemize}
Therefore, we have for $2\le \ell\le n$
\[
\E(z_{n\ell}^4)\le
\frac4{n^4}\Big(\sigma_1^{(\ell)}+4\sigma_2^{(\ell)}+3\sigma_3^{(\ell)}+6\sigma_4^{(\ell)}+\sigma_5^{(\ell)}\Big).
\]
By summing up on both sides of the above inequality we have
 \begin{eqnarray*}
\sum^n_{\ell=2}\E(z_{n\ell}^4)&\le&
\frac4{n^4}\Big(\sum^n_{\ell=2}\sigma_1^{(\ell)}+4\sum^n_{\ell=2}\sigma_2^{(\ell)}+
3\sum^n_{\ell=2}\sigma_3^{(\ell)}+6\sum^n_{\ell=2}\sigma_4^{(\ell)}+\sum^n_{\ell=2}\sigma_5^{(\ell)}\Big)\\
&\le
&(\frac{2}{n^2}\sum^{k_n}_{i=1}\frac{1}{p_i^2}+\frac{8}{n}\sum^{k_n}_{i=1}\frac{1}{p_i})+(\frac{72}{n}\sum^{k_n}_{i=1}\frac{1}{p_i}+
\frac{48k_n}{n})\\
&&+\frac{36k_n^2}{n}+\frac{72k_n}{n}+\frac{12}{n}\\
&\le &
\frac{2}{n^2}\sum^{k_n}_{i=1}\frac{1}{p_i^2}+\frac{80}{n}\sum^{k_n}_{i=1}\frac{1}{p_i}
+\frac{36(k_n+2)^2}{n}.
\end{eqnarray*}
Since $\sigma_{n1}^2\sim 2k_n$, equation \eqref{4thmoment} follows
immediately from \eqref{sonofC3} and condition \eqref{C3}. This
completes the proof. \qed

\begin{lemma}\label{lem4} Let $c_i\ge 0$, $1\le i\le k_n$, be given weights such that $\sum^{k_n}_{i=1}c_i=k_n$.
 Assume $\{n_r,~ r\ge 1\}$ is an increasing sequence of positive integers.
 If
\begin{equation}\label{clt-link}
 \frac{\beta_{n_r3}}{n_r\beta_{n_r1}^2}\to 0 ~~\mbox{ and  }~~\frac{k_{n_r}^2}{n_r\beta_{n_r1}}\to 0 \mbox{ as }r\to\infty,
\end{equation}
where $\beta_{n_rj}$'s are defined in \eqref{jj}, then we have
\begin{equation}\label{CLT3}
\frac{\overline{S}_{n_r2}}{\overline\sigma_{n_r2}}\td
N(0,1)~~~\mbox{ as } r\to\infty,
\end{equation}
where $\overline\sigma_{n2}$ is defined in \eqref{sigma2bar}. If,
additionally, \eqref{C3} holds with $n=n_r$ as $r\to\infty$, we have
\begin{equation}\label{indepdendence}
\Big(\frac{S_{n_r1}-(k_{n_r}-1)}{\sigma_{n_r1}},
\frac{\overline{S}_{n_r2}}{\overline\sigma_{n_r2}}\Big)\td
(Z_1,Z_2),
\end{equation}
where $Z_1$ and $Z_2$ are i.i.d. standard normal random variables.
\end{lemma}

\noindent{\it Proof}. As in the proof of Lemma~\ref{lem3}, we denote
$n_r$ as $n$ for brevity.

It follows from \eqref{S2bar} and \eqref{o-e} that
\begin{eqnarray*}
\overline{S}_{n2}&=&\sum^{k_n}_{i=1}c_i\Big(\frac{o_i}{e_i}-1\Big)\\
&=&\frac1{n}\sum^{k_n}_{i=1}
\Big(\frac{c_i}{p_i}\sum^n_{j=1}I(X_j=i)-c_in\Big)\\
&=&\frac{1}{n}\sum^n_{j=1}\Big(\sum^{k_n}_{i=1}\frac{c_iI(X_j=i)}{p_i}-k_n\Big).
\end{eqnarray*}
Set
\begin{equation}\label{ynj}
y_{nj}=\sum^{k_n}_{i=1}\frac{c_iI(X_j=i)}{p_i}-k_n,~~~1\le j\le n.
\end{equation}
Then $\overline{S}_{n2}=\frac1n\sum^n_{j=1}y_{nj}$. Note that
$y_{n1}, \cdots, y_{nn}$ are $n$ i.i.d. random variables with mean
$0$. Since for any integer $r\ge 2$
\[
\E\Big(\sum^{k_n}_{i=1}\frac{c_iI(X_j=i)}{p_i}\Big)^r=\E\Big(\sum^{k_n}_{i=1}\frac{c_i^rI(X_j=i)}{p_i^r}\Big)=\sum^{k_n}_{i=1}\frac{c_i^r}{p_i^{r-1}},
\]
we have
\[
\E(y_{n1}^2)=\E\Big(\sum^{k_n}_{i=1}\frac{c_iI(X_j=i)}{p_i}\Big)^2-k_n^2
=\sum^{k_n}_{i=1}\frac{c_i^2}{p_i}-k_n^2=\beta_{n1},
\]
which implies
\begin{equation}\label{varofS2}
\overline\sigma_{n2}^2=\mathrm{Var}(\overline{S}_{n2})=\frac{\beta_{n1}}{n}.
\end{equation}
Furthermore, we have
\begin{eqnarray*}
\E(y_{n1}^4)&=&\E\Big(\sum^{k_n}_{i=1}\frac{c_iI(X_j=i)}{p_i}-k_n\Big)^4\\
&=&\E\Big(\sum^{k_n}_{i=1}\frac{c_iI(X_j=i)}{p_i}\Big)^4-4k_n\E\Big(\sum^{k_n}_{i=1}\frac{c_iI(X_j=i)}{p_i}\Big)^3\\
&&~~+6k_n^2\E\Big(\sum^{k_n}_{i=1}\frac{c_iI(X_j=i)}{p_i}\Big)^2-4k_n^3\E\Big(\sum^{k_n}_{i=1}\frac{c_iI(X_j=i)}{p_i}\Big)+k_n^4\\
&=&\sum^{k_n}_{i=1}\frac{c_i^4}{p_i^3}-k_n^4-4k_n\Big(\sum^{k_n}_{i=1}\frac{c_i^3}{p_i^2}-k_n^3\Big)
+6k_n^2\Big(\sum^{k_n}_{i=1}\frac{c_i^2}{p_i}-k_n^2\Big)\\
&\le&\sum^{k_n}_{i=1}\frac{c_i^4}{p_i^3}-k_n^4+6k_n^2\Big(\sum^{k_n}_{i=1}\frac{c_i^2}{p_i}-k_n^2\Big)\\
&=&\beta_{n3}+6k_n^2\beta_{n1}
\end{eqnarray*}
from \eqref{jj}.

Note that as $n\to\infty$
\begin{equation}\label{lyapunov}
\frac{1}{(\sqrt{n\beta_{n1}})^4}\sum^{n}_{j=1}\E(y_{nj}^4)=\frac{\E(y_{n1}^4)}{n\beta_{n1}^2}\le
\frac{\beta_{n3}}{n\beta_{n1}^2}+\frac{6k_n^2}{n\beta_{n1}}\to 0
\end{equation}
from \eqref{clt-link}.
 This is Lyapunov's condition for the central limit theorem
\[
\frac{\overline{S}_{n2}}{\overline\sigma_{n2}}=\frac{\sum^{n}_{j=1}y_{nj}}{\sqrt{n\beta_{n1}}}\td
N(0,1).
\]
Therefore, we have proved \eqref{CLT3}.

Since both $\frac{S_{n1}-(k_n-1)}{\sigma_{n1}}$ and
$\frac{\overline{S}_{n2}}{\overline\sigma_{n2}}$ converge in
distribution to the standard normal, to show \eqref{indepdendence},
it suffices to show that for any $s,t\in \mathbb{R}$
\[
s\frac{S_{n1}-(k_n-1)}{\sigma_{n1}}+t\frac{\overline{S}_{n2}}{\overline\sigma_{n2}}\td
N(0, s^2+t^2),
\]
or equivalently
\begin{equation}\label{last-clt}
T_n(s,t):=\frac{s}{\sqrt{s^2+t^2}}\frac{S_{n1}-(k_n-1)}{\sigma_{n1}}+\frac{t}{\sqrt{s^2+t^2}}\frac{\overline{S}_{n2}}{\overline\sigma_{n2}}\td
N(0, 1).
\end{equation}

Now fix $s, t\in \mathbb{R}$. Set
\[
a_n=\frac{s}{\sqrt{s^2+t^2}}\frac{1}{\sigma_{n1}},
~~b_n=\frac{s}{\sqrt{s^2+t^2}}\frac{1}{\sqrt{n\beta_{n1}}}.
\]
Note that
$y_{n\ell}=\sum^{k_n}_{i=1}\frac{c_i\delta_{i,\ell}}{p_i}$. Define
\[
x_{n\ell}=a_nz_{n\ell}+b_ny_{n\ell},
\]
where $z_{n\ell}$'s are defined in \eqref{znl} in the proof of
Lemma~\ref{lem3}. Then we have
\[
T_n(s,t)=\sum^n_{\ell=1}x_{n\ell}.
\]
Obviously, $\{x_{n\ell}, \mathcal{F}_{n\ell}, ~1\le \ell \le n,
~n\ge 1\}$ is an array of martingale differences.

In view of \eqref{fact3} and \eqref{fact4}, we have
$\E(\delta_{i,\ell}\delta_{i',\ell})=p_iI(i=i')-p_ip_{i'}$ and
$\sum^{k_n}_{i=1}\Delta_{i,\ell-1}=0$, which imply
\begin{eqnarray*}
\E(y_{n\ell}z_{n\ell}|\mathcal{F}_{n(\ell-1)})
&=&\frac{2}{n}\sum^{k_n}_{i=1}\sum^{k_n}_{i'=1}\frac{c_{i'}}{p_ip_{i'}}\Delta_{i,\ell-1}\E(\delta_{i,\ell}\delta_{i',\ell})\\
&=&\frac{2}{n}\sum_{1\le i=i'\le
k_n}\frac{c_{i'}}{p_ip_{i'}}\Delta_{i,\ell-1}p_i
-\frac{2}{n}\sum^{k_n}_{i=1}\sum^{k_n}_{i'=1}\frac{c_{i'}}{p_ip_{i'}}\Delta_{i,\ell-1}p_ip_{i'}\\
&=&\frac{2}{n}\sum^{k_n}_{i=1}\frac{c_i\Delta_{i,\ell-1}}{p_i}
-\frac{2}{n}\sum^{k_n}_{i'=1}c_{i'}\sum^{k_n}_{i=1}\Delta_{i,\ell-1}\\
&=&\frac{2}{n}\sum^{k_n}_{i=1}\frac{c_i\Delta_{i,\ell-1}}{p_i}.
\end{eqnarray*}
Therefore, we have
\[
\E\big(\E(y_{n\ell}z_{n\ell}|\mathcal{F}_{n(\ell-1)})\big)=0.
\]
Define
\[
\tau_n=\sum^{n}_{\ell=1}\E(y_{n\ell}z_{n\ell}|\mathcal{F}_{n(\ell-1)}).
\]
Then $\tau_n$ can be written as
\[
\tau_n=\frac{2}{n}\sum^{n-1}_{j=1}(n-j)\sum^{k_n}_{i=1}\frac{c_i\delta_{i,j}}{p_i}=\frac{2}{n}\sum^{n-1}_{j=1}(n-j)y_{nj},
\]
where $y_{nj}$'s, as defined in \eqref{ynj}, are iid random
variables with mean $0$ and variance $\beta_{n1}$. We conclude that
\begin{equation}\label{smalltau}
\E(\tau_n^2)\le 4n\beta_{n1}.
\end{equation}

By using the formula
\[
x_{n\ell}^2=(a_nz_{n\ell}+b_ny_{n\ell})^2=a_n^2z_{n\ell}^2+b_n^2y_{n\ell}^2+2a_nb_nz_{n\ell}y_{n\ell},
\]
we get the conditional variance for the martingale differences
$\{x_{n\ell}\}$
\begin{eqnarray*}
\lambda_{n|c}^2:&=&\sum^{n}_{\ell=1}\E(x_{n\ell}^2|\mathcal{F}_{n(\ell-1)})\\
&=&a_n^2\sum^{n}_{\ell=1}\E(z_{n\ell}^2|\mathcal{F}_{n(\ell-1)})+b_n^2\sum^{n}_{\ell=1}\E(y_{n\ell}^2|\mathcal{F}_{n(\ell-1)})
+2a_nb_n\sum^n_{\ell=1}\E(z_{n\ell}y_{n\ell}|\mathcal{F}_{n(\ell-1)})\\
&=&a_n^2\sigma_{n|c}^2+nb_n^2\beta_{n1}+2a_nb_n\tau_n.
\end{eqnarray*}
Therefore, we have from \eqref{sigman|c} and \eqref{Esigman|c} that
\[
\E(\lambda_{n|c}^2)=a_n^2\sigma_{n1}^2+nb_n^2\beta_{n1}+0=1.
\]
By using the same argument as that in the proof of Lemma~\ref{lem3},
if we can show
\begin{equation}\label{4thmomentx}
\sum^n_{\ell=1}\E(x_{n\ell}^4)=o(1)
\end{equation}
and
\begin{equation}\label{cvariancex}
\E(\lambda_{n|c}^2-1)^2=o(1),
\end{equation}
then we can apply the martingale central limit theorem to obtain
\eqref{last-clt}. In fact, since
\[
\lambda_{n|c}^2-1=a_n^2(\sigma_{n|c}^2-\sigma_{n1}^2)+2a_nb_n\tau_n,
\]
 we have from the $c_r$-inequality
 \begin{eqnarray*}
 \E(\lambda_{n|c}^2-1)^2&\le& 2\Big(a_n^4\E(\sigma_{n|c}^2-\sigma_{n1}^2)^2+4a_n^2b_n^2\E(\tau_n^2)\Big)\\
 &\le &2\Big(\frac{\E(\sigma_{n|c}^2-\sigma_{n1}^2)^2}{\sigma_{n1}^4}+\frac{4\E(\tau_n^2)}{\sigma_{n1}^2n\beta_{n1}}\Big)\\
 &\to& 0
 \end{eqnarray*}
in view of  \eqref{2ndmoment} and \eqref{smalltau}. This proves
\eqref{cvariancex}.  Again, by using the $c_r$-inequality we have
that as $n\to\infty$
\begin{eqnarray*}
\sum^n_{\ell=1}\E(x_{n\ell}^4) &\le& 8\sum^n_{\ell=1}\Big(\E(a_n^4z_{n\ell}^4)+b_n^4\E(y_{n\ell}^4)\Big)\\
&\le &\frac{8}{\sigma_{n1}^4}\sum^n_{\ell=1}\E(z_{n\ell}^4)+\frac{1}{(\sqrt{n\beta_{n1}})^4}\sum^n_{\ell=1}\E(y_{n\ell}^4)\\
&\to&0
\end{eqnarray*}
from \eqref{4thmoment} and \eqref{lyapunov}, proving
\eqref{4thmomentx}. This completes the proof of the lemma. \qed

\vspace{10pt} \noindent{\it Proof of Theorem~\ref{thm1}.}
Theorem~\ref{thm1} is a direct consequence of Lemma~\ref{lem3} when
$n_r$ is the entire sequence of all positive integers. \qed

\vspace{10pt}

\noindent{\it Proof of Theorem~\ref{thm2}.} We will employ
subsequence arguments, that is, \eqref{CLT} holds if and only if for
any increasing sequence of positive integers, there exists its
further subsequence, say, $\{n_r, ~r\ge 1\}$ such that \eqref{CLT}
holds along $n=n_r$ as $r\to\infty$.  Since
$\sigma_{n2}^2/\sigma_{n}^2\in [0,1)$,  $n_r$ can be selected in a
way that  $\sigma_{n_r2}^2/\sigma_{n_r}^2$ has a limit in $[0,1]$.
Therefore, it suffices to show that \eqref{CLT} holds for $n=n_r$
for any increasing sequence of integers $\{n_r\}$ as long as
\[
\lim_{r\to\infty}\frac{\sigma_{n_r2}^2}{\sigma_{n_r1}^2}=v ~\mbox{
for some }v\in [0,1].
\]

First, consider the case  $v=0$. From Chebyshev's inequality, we
have that for every $\delta>0$
\[
P(\frac{|S_{n_r2}|}{\sigma_{n_r}}>\delta)\le
\frac{1}{\delta^2}\E\Big(\frac{|S_{n_r2}|}{\sigma_{n_r}}\Big)^2
=\frac{1}{\delta^2}\frac{\sigma_{n_r2}^2}{\sigma_{n_r}^2}\to 0
\] as $r\to\infty$. This implies $S_{n_r2}/\sigma_{n_r}$ converges
to zero in probability as $r\to\infty$. Since \eqref{C3} implies
\eqref{CLT2} from Theorem~\ref{thm1}, we obtain
\begin{eqnarray*}
\frac{\mathcal{X}^2_{n_r}-(k_{n_r}-1)}{\sigma_{n_r}}&=&\frac{\sigma_{n_r1}}{\sigma_{n_r}}\frac{S_{n_r1}-(k_{n_r}-1)}{\sigma_{n_r1}}+
\frac{S_{n_r2}}{\sigma_{n_r}}\\
&=&(1+o(1))\frac{S_{n_r1}-(k_{n_r}-1)}{\sigma_{n_r1}}+o_p(1)\\
&\td& N(0,1)
\end{eqnarray*}
as $r\to\infty$, i.e. \eqref{CLT} holds with $n=n_r$.

Now consider the case $v\in (0,1]$. We can use Lemma~\ref{lem4} for
$c_1=\cdots=c_{k_n}=1$. In this case, $\overline{S}_{n2}=S_{n2}$,
$\overline\sigma_{n2}=\sigma_{n2}$, and conditions \eqref{C44} and
\eqref{C4} are the same.

Since $\sigma_{n1}^2\sim 2k_n$, we have
\[
\frac{\sigma_{n_r2}^2}{k_{n_r}}\sim
\frac{2\sigma_{n_r2}^2}{\sigma_{n_r1}^2}=\frac{2\sigma_{n_r2}^2}{\sigma_{n_r}^2-\sigma_{n_r2}^2}
\to \frac{2v}{1-v}>0.
\]
The above limit is interpreted as infinity if $v=1$. This, together
with \eqref{C4}, implies that the first term within the parentheses
in \eqref{C4} must tend to zero as $n=n_r$ goes to infinity, that
is,
\[
\frac{\beta_{n_r3}}{n_r\beta_{n_r1}^2}=\frac{\sum^{k_{n_r}}_{i=1}\frac{1}{p_i^3}-k_{n_r}^4}{n_r^3\sigma_{n_r2}^4}\to
0
\]
as $r\to\infty$. We have used \eqref{varofS2} here. Furthermore,
\eqref{C3} and \eqref{jj} with $j=2$ imply that $\frac{k_n}{n^2}\to
0$ as $n\to\infty$, and thus
\[
\frac{k_{n_r}^2}{n_r\beta_{n_r1}}=\frac{k_{n_r}}{n_r^2}\frac{k_{n_r}}{\sigma_{n_r1}^2}\to
0~~\mbox{ as }r\to\infty.
\]
Therefore, \eqref{clt-link} is satisfied.  In view of
\eqref{indepdendence} we have
\[
\frac{\mathcal{X}^2_{n_r}-(k_{n_r}-1)}{\sigma_{n_r}}=\frac{\sigma_{n_r1}}{\sigma_{n_r}}\frac{S_{n_r1}-(k_{n_r}-1)}{\sigma_{n_r1}}+
\frac{\sigma_{n_r2}}{\sigma_{n_r}}\frac{S_{n_r2}}{\sigma_{n_r2}}\td
\sqrt{v}Z_1+\sqrt{1-v}Z_2.
\]
The above limit is a standard normal random variable. Thus, we have
proved \eqref{CLT} with $n=n_r$.  \qed

\vspace{10pt} \noindent{\it Proof of Theorem~\ref{thm3}.}
Theorem~\ref{thm3} is a special case of Theorem~\ref{thm4}. \qed

\vspace{10pt} \noindent{\it Proof of Theorem~\ref{thm4}.} When
$c=0$, the test statistic $S_{n1}+c|\overline{S}_{n2}|$ is the same
as $S_{n1}$, and Theorem~\ref{thm1} ensures Theorem~\ref{thm4}.
Therefore, we focus on the case $c>0$. We note that
\begin{eqnarray*}
& &\sup_{x}\Big|P\big(\frac{S_{n1}+c|\overline{S}_{n2}|}{\sigma_{n1}}\le x\Big)-P\Big(Z_1+\frac{c\overline\sigma_{n2}}{\sigma_{n1}}|Z_2|\le x\Big)\Big|\\
&=&\sup_{x}\Big|P\Big(\frac{1}{1+\frac{c\overline\sigma_{n2}}{\sigma_{n1}}}\frac{S_{n1}+c|\overline{S}_{n2}|}{\sigma_{n1}}\le
x\Big)-P\Big(\frac{1}{1+\frac{c\overline\sigma_{n2}}{\sigma_{n1}}}(Z_1+\frac{c\overline\sigma_{n2}}{\sigma_{n1}}|Z_2|)\le
x\Big)\Big|=:\Theta_n.
\end{eqnarray*}
We will also use subsequence arguments as those in the proof of
Theorem~\ref{thm2}. To show that $\Theta_n$ converges to zero, it
suffices to prove that $\Theta_{n_r}\to 0$ as $r\to\infty$ for every
increasing sequence of integers $\{n_r\}$ such that
\[
\displaystyle\frac{\frac{c\overline\sigma_{n_r2}}{\sigma_{n_r1}}}{1+\frac{c\overline\sigma_{n_r2}}{\sigma_{n_r1}}}\to
v~~\mbox{ for some } v\in [0,1].
\]
The proof is similar to that in the proof of Theorem~\ref{thm2}.

When $v=0$, we have
$\frac{c\overline\sigma_{n_r2}}{\sigma_{n_r1}}\to 0$ as
$r\to\infty$. By using Chebyshev's inequality we can show that
\[
\frac{c\overline{S}_{n_r2}}{\sigma_{n_r1}}=\frac{c\overline\sigma_{n_r2}}{\sigma_{n_r1}}\frac{\overline{S}_{n_r2}}{\overline\sigma_{n_r2}}~~\mbox{
converges to zero in probability},
\]
which, coupled with Lemma~\ref{lem3}, yields that
\[
\frac{1}{1+\frac{c\overline\sigma_{n_r2}}{\sigma_{n_r1}}}\frac{S_{n_r1}+c|\overline{S}_{n_r2}|}{\sigma_{n_r1}}
=\frac{1}{1+\frac{c\overline\sigma_{n_r2}}{\sigma_{n_r1}}}\frac{S_{n_r1}}{\sigma_{n_r1}}+\frac{c\overline{S}_{n_r2}}{\sigma_{n_r1}}
=(1+o(1))\frac{S_{n_r1}}{\sigma_{n_r1}}+o_p(1)\td N(0,1).
\]
Obviously, we have
$\frac{1}{1+\frac{c\overline\sigma_{n_r2}}{\sigma_{n_r1}}}(Z_1+\frac{c\overline\sigma_{n_r2}}{\sigma_{n_r1}}|Z_2|)\to
Z_1$. Therefore, we get
\begin{equation}\label{limit1}
\sup_x|P\Big(\frac{1}{1+\frac{c\overline\sigma_{n_r2}}{\sigma_{n_r1}}}\frac{S_{n_r1}+c|\overline{S}_{n_r2}|}{\sigma_{n_r1}}\le
x\Big)-\Phi(x)|\to 0
\end{equation}
and
\begin{equation}\label{limit2}
\sup_x|P\Big(\frac{1}{1+\frac{c\overline\sigma_{n_r2}}{\sigma_{n_r1}}}(Z_1+\frac{c\overline\sigma_{n_r2}}{\sigma_{n_r1}}|Z_2|)\le
x\Big)-\Phi(x)|\to 0
\end{equation}
as $r\to\infty$. By using the triangle inequality, $\Theta_{n_r}$ is
dominated by the sum of the two suprema above and thus converges to
zero.

When $v\in (0,1]$, by following the same arguments in the proof of
Theorem~\ref{thm2}, we can show  \eqref{clt-link} is satisfied.
Hence, we can have \eqref{indepdendence}, and both
$\frac{1}{1+\frac{c\overline\sigma_{n_r2}}{\sigma_{n_r1}}}\frac{S_{n_r1}+c|\overline{S}_{n_r2}|}{\sigma_{n_r1}}$
and
$\frac{1}{1+\frac{c\overline\sigma_{n_r2}}{\sigma_{n_r1}}}(Z_1+\frac{c\overline\sigma_{n_r2}}{\sigma_{n_r1}}|Z_2|)$
converge in distribution to $(1-v)Z_1+v|Z_2|$ which is a continuous
random variable. Denote the cumulative distribution of this limit as
$\Phi_v$.   Then \eqref{limit1} and \eqref{limit2} hold if $\Phi$ is
replaced by $\Phi_v$. Again, by using the triangle inequality we get
that $\Theta_{n_r}$ converges to zero as $r\to\infty$. \qed


\begin{thebibliography}{99}
\bibitem{Pearson}
Pearson, K. (1900). On the criterion that a given system of deviations from the probable in the case
of a correlated system of variables is such that it can be reasonably supposed to have arisen from
random sampling, {\it Philosophical Magazine} \textbf{50}, 157--175.
\end{thebibliography}
\end{document}